\documentclass[twocolumn,tighten]{aastex63}
\usepackage{graphicx}
\usepackage{url}
\usepackage{amsmath,amsfonts}
\usepackage{apjfonts}
\usepackage{needspace}
\usepackage{hyperref}
\usepackage[figure,figure*]{hypcap}

\newcommand{\T}[1]{\ensuremath{T_\mathrm{#1}}}
\newcommand{\calT}[1]{\ensuremath{\mathcal{T}_\mathrm{#1}}}
\newcommand{\E}[1]{\ensuremath{\langle #1 \rangle}}
\newcommand{\cc}[2]{\ensuremath{#1^{\vphantom{\ast}} #2^{\ast}}}
\newcommand{\Arg}[1]{\ensuremath{\mathop{\mathrm{Arg}}[#1]}}

\newcommand{\nrao}{National Radio Astronomy Observatory, 520 Edgemont Rd, Charlottesville, VA 22903, USA}
\newcommand{\naoj}{National Astronomical Observatory of Japan, 2-21-1 Osawa, Mitaka, Tokyo 181-8588, Japan}
\newcommand{\haystack}{Massachusetts Institute of Technology, Haystack Observatory, 99 Millstone Rd, Westford, MA 01886, USA}
\newcommand{\bhi}{Black Hole Initiative, Harvard University, 20 Garden Street, Cambridge, MA 02138, USA}
\newcommand{\radboud}{Department of Astrophysics/IMAPP, Radboud University, P.O. Box 9010, 6500 GL Nijmegen, The Netherlands}
\newcommand{\cfa}{Center for Astrophysics $|$ Harvard \& Smithsonian, 60 Garden Street, Cambridge, MA 02138, USA}
\newcommand{\arizona}{University of Arizona, 933 North Cherry Avenue, Tucson, AZ 85721, USA}
\newcommand{\cit}{California Institute of Technology, 1200 East California Boulevard, Pasadena, CA 91125, USA}

\newcommand{\mw}[1]{\textcolor{teal}{MW: #1}}
\newcommand{\ck}[1]{\textcolor{orange}{CK: #1}}
\newcommand{\lb}[1]{\textcolor{blue}{LB: #1}}
\newcommand{\mj}[1]{\textcolor{orange}{MJ: #1}}
\newcommand{\mdj}[1]{\textcolor{olive}{MDJ: #1}}
\newcommand{\si}[1]{\textcolor{magenta}{SI: #1}}

\renewcommand{\mw}[1]{}
\renewcommand{\ck}[1]{}
\renewcommand{\lb}[1]{}
\renewcommand{\mj}[1]{}
\renewcommand{\mdj}[1]{}
\renewcommand{\si}[1]{}

\begin{document}

\def\sgra{Sgr\,A$^{\ast}$}
\def\lsim{\mathrel{\raise.3ex\hbox{$<$\kern-.75em\lower1ex\hbox{$\sim$}}}}
\def\gsim{\mathrel{\raise.3ex\hbox{$>$\kern-.75em\lower1ex\hbox{$\sim$}}}}

\title{EHT-HOPS pipeline for millimeter VLBI data reduction}

\author{Lindy Blackburn}
\affil{\cfa}
\affil{\bhi}
\author{Chi-kwan Chan}
\affil{\arizona}
\author{Geoffrey B. Crew}
\affil{\haystack}
\author{Vincent L. Fish}
\affil{\haystack}
\author{Sara Issaoun}
\affil{\radboud}
\author{Michael D. Johnson}
\affil{\cfa}
\affil{\bhi}
\author{Maciek Wielgus}
\affil{\cfa}
\affil{\bhi}
\author{Kazunori Akiyama}
\affil{\nrao}
\affil{\haystack}
\affil{\naoj}
\author{John Barrett}
\affil{\haystack}
\author{Katherine L. Bouman}
\affil{\cit}
\affil{\cfa}
\affil{\bhi}
\author{Roger Cappallo}
\affil{\haystack}
\author{Andrew A. Chael}
\affil{\cfa}
\affil{\bhi}
\author{Michael Janssen}
\affil{\radboud}
\author{Colin J. Lonsdale}
\affil{\haystack}
\author{Sheperd S. Doeleman}
\affil{\cfa}
\affil{\bhi}

\begin{abstract}
We present the design and implementation of an automated data calibration and reduction pipeline for very long baseline interferometric (VLBI) observations taken at millimeter wavelengths.
These short radio wavelengths provide the best imaging resolution available from ground-based VLBI networks such as the Event Horizon Telescope (EHT) and the Global Millimeter VLBI Array (GMVA) but require specialized processing owing to the strong effects from atmospheric opacity and turbulence as well as the heterogeneous nature of existing global arrays. The pipeline builds on a calibration suite (HOPS)
originally designed for precision geodetic VLBI. To support the reduction of data for astronomical observations, we have developed an additional framework for global phase and amplitude calibration that provides output in a standard data format for astronomical imaging and analysis.
The pipeline was successfully used toward the reduction of 1.3\,mm observations from the EHT 2017 campaign, leading to the first image of a black hole ``shadow" at the center of the radio galaxy M87.
In this work, we analyze observations taken at 3.5\,mm (86\,GHz) by the GMVA, joined by the phased Atacama Large Millimeter/submillimeter Array in 2017 April, and demonstrate the benefits from the specialized processing of high-frequency VLBI data with respect to classical analysis techniques.
\end{abstract}


\section{Introduction}

In the technique of very long baseline interferometry (VLBI), signals from an astronomical source are recorded independently at multiple locations and later brought together for pairwise correlation. This process samples the coherence function of the incident radiation at separations corresponding to the baseline vectors between sites. The resolution probed by a baseline is determined by the interferometric fringe spacing, $1/|\mathbf{u}| = \lambda/D_\mathrm{proj}$, in angular units on the sky, where the two-dimensional spatial frequency $\mathbf{u} = (u,v)$ corresponds to the projected baseline vector in units of observing wavelength $\lambda$. Thus, the highest resolutions are achieved when sites have the widest possible separation $D$ and observe at the highest possible frequencies $\nu = c/\lambda$.

Two global networks exist for millimeter VLBI observations. The Global Millimeter VLBI Array\footnote{\url{https://www3.mpifr-bonn.mpg.de/div/vlbi/globalmm}} (GMVA) operates at 3.5\,mm (86\,GHz) and includes the Very Long Baseline Array\footnote{\url{https://science.nrao.edu/facilities/vlba}} (VLBA) and a number of large-aperture dishes with the required surface accuracy and sufficiently good local weather to operate at 3.5\,mm. The Event Horizon Telescope\footnote{\url{https://eventhorizontelescope.org}} (EHT) operates as an array at 1.3\,mm (230\,GHz), a wavelength at which only a handful of existing sites globally are able to observe. In 2017 April,
both networks participated in science observations
for the first time with
the Atacama Large Millimeter/submillimeter Array (ALMA). ALMA acted as a phased array of $\sim$37 dishes \citep{matthews2018,Goddi_2019}, providing a highly sensitive anchor station that greatly expanded the sensitivity, resolution, and baseline coverage of the VLBI networks.
In particular, the EHT 2017 array, operating over six geographical locations and including ALMA, was able to reach the necessary sensitivity and coverage in order to form the first VLBI images reconstructed at 1.3\,mm wavelength and the necessary resolution in order to image and characterize the horizon-scale supermassive black hole ``shadow" at the center of the radio galaxy M87 \citep{PaperI,PaperII,PaperIII,PaperIV,PaperV,PaperVI}.

At the heart of the VLBI technique is the correlation of the raw
station data using either dedicated hardware or software.
The correlation is
manifest as an interference fringe that changes in an expected way
as the Earth rotates. This is a simple but computationally
expensive process that requires good, but nevertheless approximate,
models in order to measure the interferometric
fringe. Some
post-correlation processing is then required to detect and analyze the
fringes to obtain scientifically useful results.

The VLBI correlator estimates the complex correlation for signals $x_1$ and $x_2$ between pairs of antennas,
\begin{equation}
r_{12} = \frac{\E{\cc{x_1}{x_2}}}{\eta_Q\,\sqrt{ \E{\cc{x_1}{x_1}}\E{\cc{x_2}{x_2}}}} = \frac{e^{i\theta_1}\,e^{-i\theta_2}\,\mathcal{V}_{12}}{\sqrt{\mathrm{SEFD}_1 \times \mathrm{SEFD}_2}}.
\label{eqn:rij}
\end{equation}
In this expression, $\eta_Q$ is a correction factor of $\sim$0.88 accounting for the introduction of quantization noise during 2-bit digitization and $\mathcal{V}$ ($\sim$1\,Jy for bright continuum sources) is the correlated flux density that varies by baseline. The {\em system-equivalent flux density} (SEFD $\sim 10^4$\,Jy, see \autoref{sec:apriori}) reflects the original analog system noise $\eta_Q \E{\cc{x}{x}}$ in effective flux units of an astronomical source above the atmosphere, and the $e^{i\theta}$ are station phase terms corresponding to residual geometric, atmospheric, and instrumental phase suffered by the signal before it is recorded. We adopt the convention of \citet{rogers1974} where positive delay (and unwrapped phase) corresponds to the signal arriving at station 2 after station 1.

The primary residual systematics after correlation are small errors in {\em delay} and {\em delay rate}, which are related to the first-order variation of the baseline phase, \mbox{$\phi = \mathrm{Arg}[r]$}, of the complex correlation between two sites in time and frequency ($t, \nu)$:
\begin{equation}
\Delta \phi = \frac{\partial\phi}{\partial \nu} \Delta \nu + \frac{\partial\phi}{\partial t} \Delta t.
\end{equation}
Since phase error $\phi = 2\pi \nu \tau$, the delay and delay rate are given by \citep[A12.28, A12.22]{tms}
\begin{equation}
\label{eqn:delayandrate}
\tau = \frac{1}{2\pi} \frac{\partial\phi}{\partial \nu} \qquad
\dot{\tau} = \frac{1}{2\pi\nu} \frac{\partial\phi}{\partial t}.
\end{equation}
For linear phase drift, the coherence has a sinc profile
\begin{equation}
\label{eqn:fringe}
\frac{1}{\Delta\phi} \int_{-\Delta\phi/2}^{\Delta\phi/2} d\phi\,\cos\phi = \frac{\sin{(\Delta\phi/2)}}{\Delta\phi/2},
\end{equation}
as a function of accumulated phase drift, $\Delta \phi$, so that maximum coherence occurs at the fringe solution where data are
compensated for fringe phase rotation and the accumulated $\Delta\phi \to 0$.
First-order fringe searches vary the two parameters, delay and delay rate, and search for maximum coherence in excess correlated signal power over the full bandwidth and up to the length of a scan.
The original signals are highly noise dominated ($|r| \lesssim 10^{-4}$), and generally at least the first-order fringe correction must be applied in order to coherently average a sufficient number of samples and produce a level of correlated flux above the statistical (thermal) noise.

The EHT and GMVA are composed of heterogeneous collections of individual stations with varying sensitivities and characteristics, and they target high observing frequencies over wide bandwidths.
For both VLBI networks, nonlinear phase systematics beyond the first-order fringe solution are important. These include phase variations over the observing band due to small variations in path delay versus frequency prior to digitization, as well as stochastic phase fluctuations in time due to achromatic path variations from atmospheric turbulence. The instrumental phase bandpass is typically constant over long timescales and can be solved using bright calibrator sources. Atmospheric phase is more difficult, as it is continuously varying and must be solved on-source. At millimeter wavelengths, the atmospheric phase can have a decoherence timescale of seconds, and compensating for it requires that the source be detectable on a baseline to within just some fraction of the decoherence time. The need to be able to measure and compensate for the atmosphere on-source at rapid timescales has been a primary driver of the wide recording bandwidths targeted by the EHT.

In \autoref{sec:ehthops} which follows, we introduce overall structure and algorithms behind the iterative phase calibration applied during the EHT-HOPS pipeline. In \autoref{sec:postproc}, we describe a suite of post-processing tools that perform absolute flux calibration and polarization gain ratio calibration, enabling the formation of calibrated Stokes $I$ visibility coefficients in a standard \texttt{UVFITS} file format. \autoref{sec:compute} describes the overall EHT-HOPS computing software organization and workflow. The EHT-HOPS pipeline is tested on a representative 3.5\,mm GMVA+ALMA data set in \autoref{sec:comparison}, and the output of the pipeline is compared against a classical reduction pathway for low-frequency VLBI in terms of fringe detection, consistency of measured phase and amplitude, and similarity of derived images on blazar NRAO\,530.

\section{EHT-HOPS Pipeline}
\label{sec:ehthops}

The current Haystack Observatory post-processing System\footnote{\url{https://www.haystack.mit.edu/tech/vlbi/hops.html}} (HOPS) was
born from the efforts of Alan Rogers in the late 1970s with a program
called FRNGE, which was written in FORTRAN and designed to be efficient
on an HP-21MX (later renamed HP-1000) minicomputer \citep{Rogers1970,rogers1974}.  With improvements
in hardware and software, a rewrite and augmentation of the tool set were launched in
the early 1990s by Colin Lonsdale, Roger Cappallo, and Cris Niell as
driven by the needs of the of the geodetic community and of a move to higher frequencies in astronomical VLBI.  The basic algorithms
were adopted from FRNGE, but there was a complete rewrite of the code
into (K\&R) C and substantial revisions of the input/output, control and file
structures, and graphical and summary analysis tools, resulting in the framework of the current HOPS system.
This was followed by a substantial effort in the early to mid-2000s to develop tools for optimizing signal-to-noise (S/N) and deriving correction factors
for data with imperfect coherence, based on analysis of amplitude with
coherent averaging time \citep{Rogers1995}.  Further evolution was provoked
by the reemergence of software correlation \citep[DiFX;][]{deller2011} and by the
needs of EHT-scale millimeter VLBI in the 2010s, which brings us to HOPS in its current form.

Acknowledging its geodetic heritage, HOPS was optimized for precision
on per-baseline delay and delay rate measurements, which are the fundamental quantities of interest for geodetic analysis programs.  Consequently, it is
somewhat light on support for some routine calibration processes found
in some other astronomical software packages, such as the Astronomical Image Processing System \citep[AIPS;][]{Greisen_2003} and the Common Astronomy Software Applications package \citep[CASA;][]{McMullin2007}.
Nevertheless, it provides a good framework for the reduction and
analysis of millimeter VLBI data, where the complexities of atmospheric effects
require ever more specialized processing to obtain reliable
astronomical results.

Over the past decade, HOPS has been used extensively for the analysis and reduction of early EHT data \citep[e.g.,][]{Doeleman2008,Doeleman2012,Fish2011,Fish2016,Akiyama2015,Johnson2015,Lu2018}. The HOPS suite grew to support the evolving EHT instrument, with steadily increasing bandwidth \citep{Whitney_Mark6,Vertatschitsch_2015}, dual-polarization observations \citep{Johnson2015}, and a move from the Mark4 hardware correlator \citep{Whitney_2004} to the DiFX software correlator \citep{deller2011}. Calibration strategies were also developed and implemented within HOPS in order to support the segmented averaging of amplitudes and bispectra \citep{Johnson2015,Fish2016} as well as on-source phase stabilization \citep{Johnson2015}. The techniques improved the ability to build S/N of visibility amplitude and phase information for high-frequency EHT observations, in the presence of rapid atmospheric phase fluctuations.

For the needs of the EHT campaigns of 2017 and subsequent years, we have extended the basic HOPS framework with Python-based packages, included within the EHT Analysis Toolkit\footnote{\url{https://github.com/sao-eht/eat}} (\texttt{eat}
library). The Python libraries provide a convenient Python-based interface to
the underlying HOPS binary and ASCII file formats via Python \texttt{ctypes} and Pandas
DataFrames, and they provide a community-standard \texttt{UVFITS} output data format
for downstream processing. The \texttt{eat} routines are also able to enforce a
global (station-based) calibration solution across the VLBI array, locking
together the baseline-based fringe solutions provided by the HOPS \texttt{fourfit} fringe fitter. The HOPS and \texttt{eat} software suites are
packaged together into a EHT-HOPS pipeline, with a set of driver scripts that
run an automated end-to-end calibration and reduction of EHT or GMVA
correlated data given a minimal basic configuration.

\begin{figure*}[t]
\centering 
\includegraphics[width=0.85\linewidth]{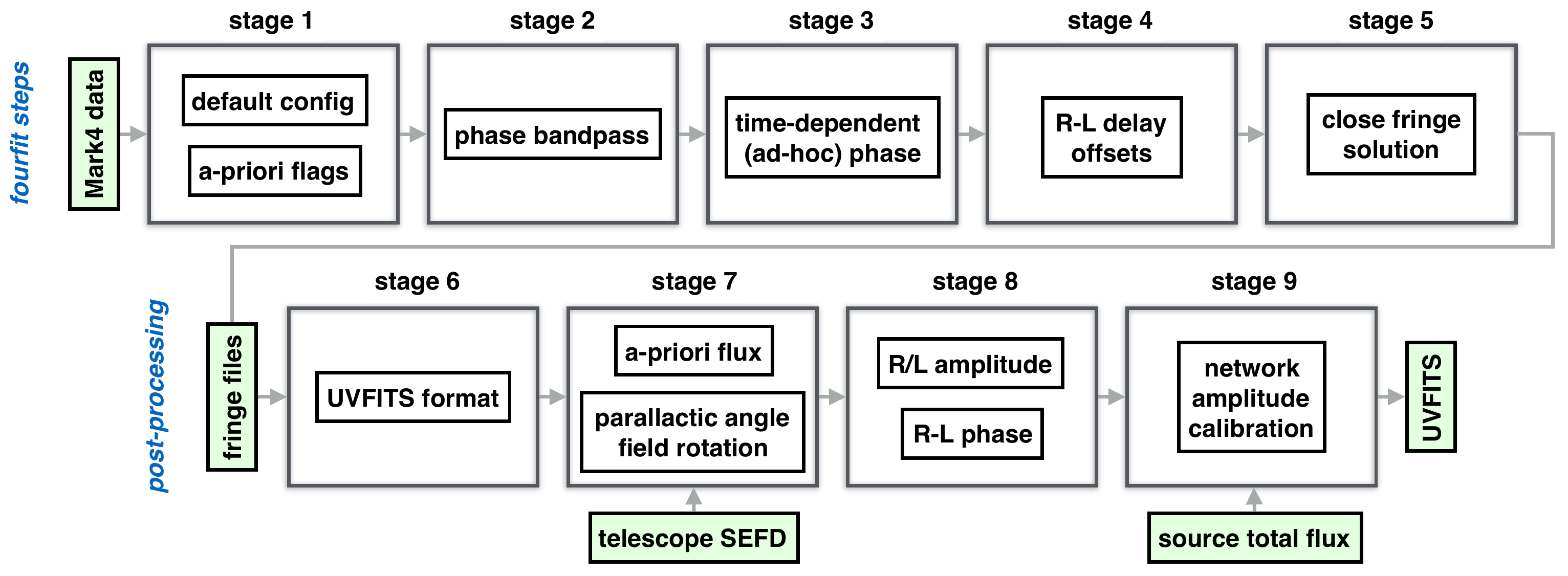}
\caption{Stages of the EHT-HOPS pipeline and post-processing steps. Stages 1--5 represent iterations of HOPS fringe fitter \texttt{fourfit}, where the input data for each stage are the original correlator output files (converted from DiFX native to Mark4 format), and the output data are a series of reduced HOPS native {\em fringe files} (averaged visibility data plus fringe solutions) and auxiliary calibration parameters (described in the inner boxes) used to refine the fringe search for successive stages. The order of the stages is not fundamental to the calibration process but is largely determined by which up-front corrections are needed to provide more precise downstream estimation of calibration parameters. After an initial run with a priori fringe search windows, channel configuration, and data flags, the residual phase bandpass and differential phase vs. time (ad-hoc phase) are calibrated to a reference station in the array during stages 2 and 3. At stage~4, precise delays are measured and aligned between RCP and LCP feeds at each station, so that a single global (station-based) fringe solution in delay and delay rate can be solved for and applied in stage~5. The output of stage~5 is converted to \texttt{UVFITS} format, and a remaining suite of post-processing tools provide amplitude calibration and time- and polarization-dependent phase calibration, as these cannot currently be performed within \texttt{fourfit}. A final stage of network calibration folds in a priori information about array redundancy and total flux density to self-calibrate colocated sites in a model-independent way.}
\label{fig:stages}
\end{figure*}

The first five stages
of the pipeline run several iterations of \texttt{fourfit} \citep{fourfitmanual},
while solving for nonlinear phase corrections and a global fringe solution.
The pipeline workflow is shown in \autoref{fig:stages},
and specific details of the \texttt{fourfit} stages are given below. Examples of various steps are provided via application of stages of the pipeline on a representative 3.5\,mm GMVA+ALMA data set from 2017 (project code MB007), the scientific results of which are published in \citet{Issaoun_2019}. Details of the observations and data reduction are given in \autoref{sect:dataset}.

\subsection{Data Flagging}

Data selection and flagging are defined using HOPS ASCII control codes. Data selection involves setting the start and stop time of processing, as well as which frequency channels are processed. Flagging defines small intervals of time within the processed segment and small frequency ranges within a channel (notches) that have their data weights set to zero and are thus ignored when fringe fitting and visibility averaging. The EHT-HOPS pipeline does not currently implement automated flagging in either time or frequency, and these must be defined by hand from data inspection and telescope logs. However, HOPS tool \texttt{aedit} and custom time series and spectral plotting tools within the \texttt {eat} library are available to assist with identifying time and frequency ranges, as well as programmatic manipulation of the relevant HOPS control codes.

\subsection{Bandpass Calibration}
\label{sec:bandpass}

\begin{figure}
    \centering
    \vspace{1em}
    \includegraphics[width=1.0\columnwidth]{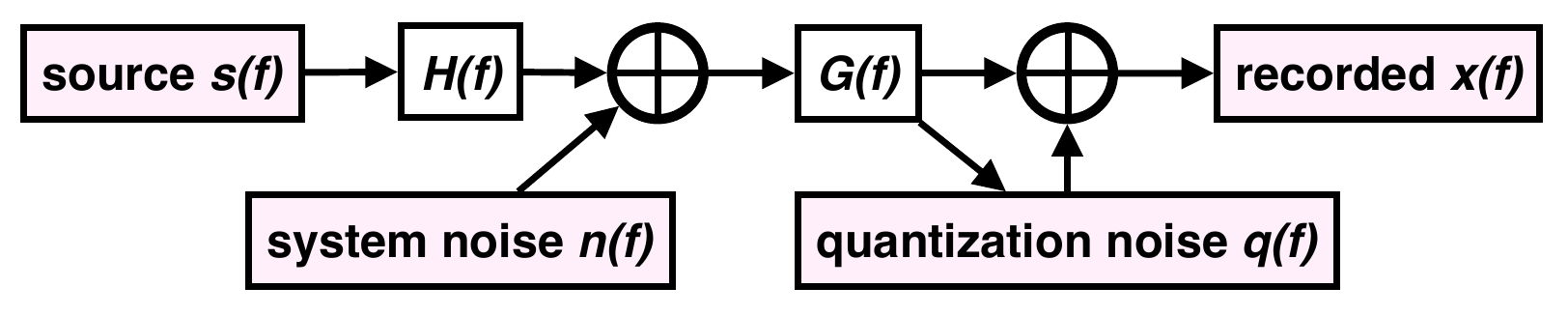}
    \caption{Simplified signal path reflecting the bandpass response of one antenna given input source $s(f)$ and system noise $n(f)$ signals represented in the frequency domain. Transfer functions $H(f)$ and $G(f)$ represent the scaling and shaping of signals as they pass through components of the environment and instrument. The recorded digitized signal $x(f) = G(Hs + n) + q$.}
    \label{fig:bandpass}
\end{figure}

\begin{figure}
    \centering
    \vspace{1em}
    \includegraphics[width=0.95\columnwidth]{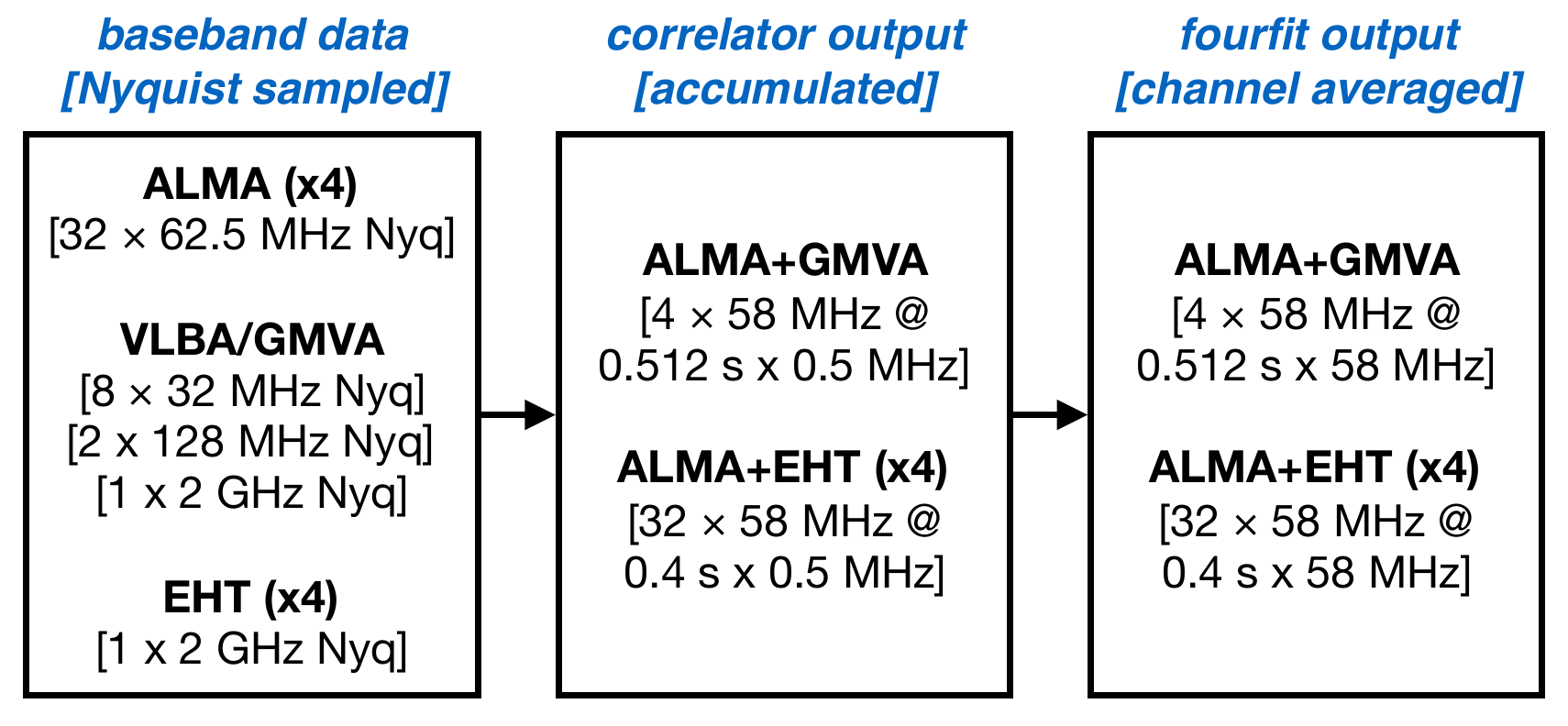}
    \caption{Time and frequency resolution of data, covering a single ALMA spectral window, as it is reduced. This represents 1/4 of the total recorded bandwidth for EHT+ALMA at each station as of 2018, where each $\sim$2\,GHz spectral window is correlated and reduced independently. Correlation parameters when ALMA is present are largely driven by the configuration of ALMA tunable filter bank (TFB) channels that are 62.5\,MHz wide, overlap slightly, and have starting frequencies aligned to 1/(32\,$\mu$s). Correlation for GMVA and EHT must therefore use an FFT window of at least 32\,$\mu$s to align to the\,MHz and currently use 64\,$\mu$s to also center the channel. The 64\,$\mu$s FFT window determines available correlation accumulation periods (APs), which must be an integer number of FFT window lengths. GMVA has chosen 0.512\,s accumulation periods, while the EHT uses 0.4\,s. Frequency accumulation is 0.5\,MHz for both networks. The raw output of HOPS fringe fitter \texttt{fourfit} maintains the original AP but averages over each 58\,MHz channel. This resolution is maintained throughout the EHT-HOPS post-processing stages, until it is time/band averaged after network calibration (not shown) for a more manageable data volume.}
    \label{fig:accumulation}
\end{figure}

Bandpass response of an antenna can be understood in context of the signal path from \autoref{fig:bandpass}. In the simplified picture, the recorded signal $x(f)$ is composed of the sum of received source signal $Hs$ and system noise $n$, subject to a common transfer function $G$ and additive quantization noise $q$ before being digitally recorded to disk: $x = G(Hs+n)+q$. $H$ includes effects such as atmospheric attenuation, dish characteristics, and receiver response. System noise $n$ includes contributions such as receiver thermal noise, atmospheric emission, and radio-frequency interference. The common transfer function $G$ accounts for components like cable transmission and back-end electronics. Finally, the effects of low-bit quantization can be approximated as additive quantization noise that depends on the signal profile prior to recording. For noise-dominated signals with a flat spectrum, quantization noise is white and uncorrelated with the source signal, and the effect on the data is modeled in a straightforward way by the correlation amplitude efficiency factor $\eta_Q$ from \autoref{eqn:rij}.

The SEFD is defined as the source flux necessary to contribute equal signal power to the system noise. In terms of elements from \autoref{fig:bandpass},
\begin{equation}
    \mathrm{SEFD} = \frac{\E{|n^2|}}{\E{|Hs|^2}}\;S
    \label{eqn:sefdbp}
\end{equation}
where $S$ is the flux density of the (unpolarized) source that generates $s$.
Ignoring quantization and assuming a noise-dominated signal, the autocorrelation spectrum of the received signal $x$ is
\begin{equation}
    \E{xx^\ast} = \E{|Gn|^2}
\end{equation}
and the cross-correlation spectrum is
\begin{equation}
    \E{x_1^{\vphantom{\ast}} x_2^\ast} = \E{s_1^{\vphantom{\ast}} s_2^\ast} \, H_1^{\vphantom{\ast}} G_1^{\vphantom{\ast}} H_2^\ast G_2^\ast
\end{equation}
as the system noise between sites is uncorrelated.

On a single baseline, the bandpass from both antennas 1 and 2 will directly affect the correlation coefficient measured (\autoref{eqn:rij}).
The DiFX software correlator \citep{deller2011}, used for both EHT and GMVA correlation, computes $\E{\cc{x_1}{x_2}}$ averaged over 1 subchannel ($\sim$0.5\,MHz) and 1 AP (accumulation period, $\sim$0.5\,s), as illustrated in \autoref{fig:accumulation}. The values for each AP are then normalized by their channel-average autocorrelation power during the DiFX$\to$Mark4 data conversion stage (using DiFX conversion tool \texttt{difx2mark4}).
This step removes the ``autocorrelation'' amplitude bandpass $|G_1^{\vphantom{\ast}} G_2^\ast|$ (at the resolution of a full channel) but leaves the residual cross-power amplitude bandpass from $|H_1^{\vphantom{\ast}} H_2^\ast/\E{n_1n_2^\ast}|$ that reflects changes in SEFD over frequency. Also left is the combined phase bandpass, $\Arg{H_1^{\vphantom{\ast}} G_1^{\vphantom{\ast}} H_2^\ast G_2^\ast} = \theta_1 - \theta_2$, which reflects very small and stable changes in instrumental path length as a function of frequency.

Stage 2 in the EHT-HOPS pipeline estimates and provides corrections for the relative {\em phase} bandpass over a baseline by averaging over an ensemble of high-S/N cross-correlation measurements to a common reference station. High-S/N fringes from the reference station (generally ALMA) to other stations in the network are taken from stage~1 output to estimate a single baseline phase and phase slope per 58\,MHz channel by direct S/N-weighted average. Baselines that do not contain the reference antenna (station 0) can then be assumed to be subject to phase bandpass $\phi_{ij} = \phi_{0j} - \phi_{0i}$.

\begin{figure}
    \centering
    \includegraphics[width=1.0\linewidth,trim=9 21 9 0,clip]{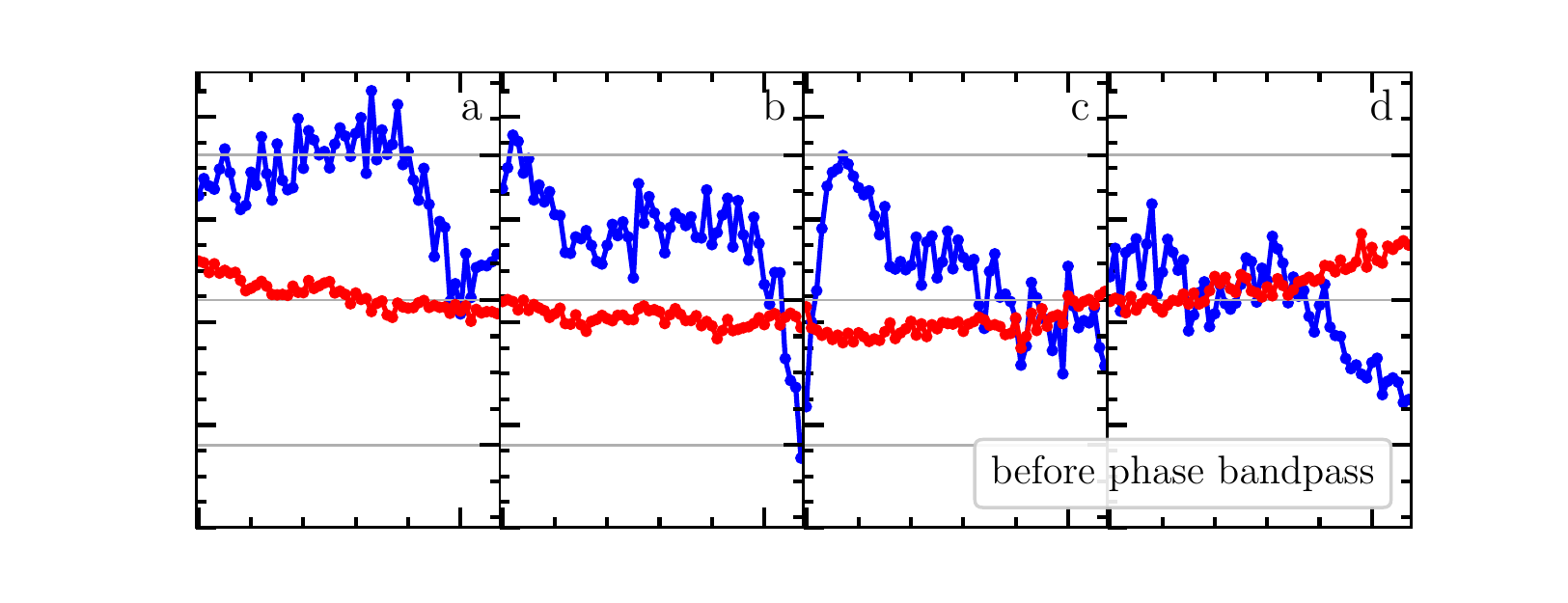} \\
    \includegraphics[width=1.0\linewidth,trim=9 0 9 21,clip]{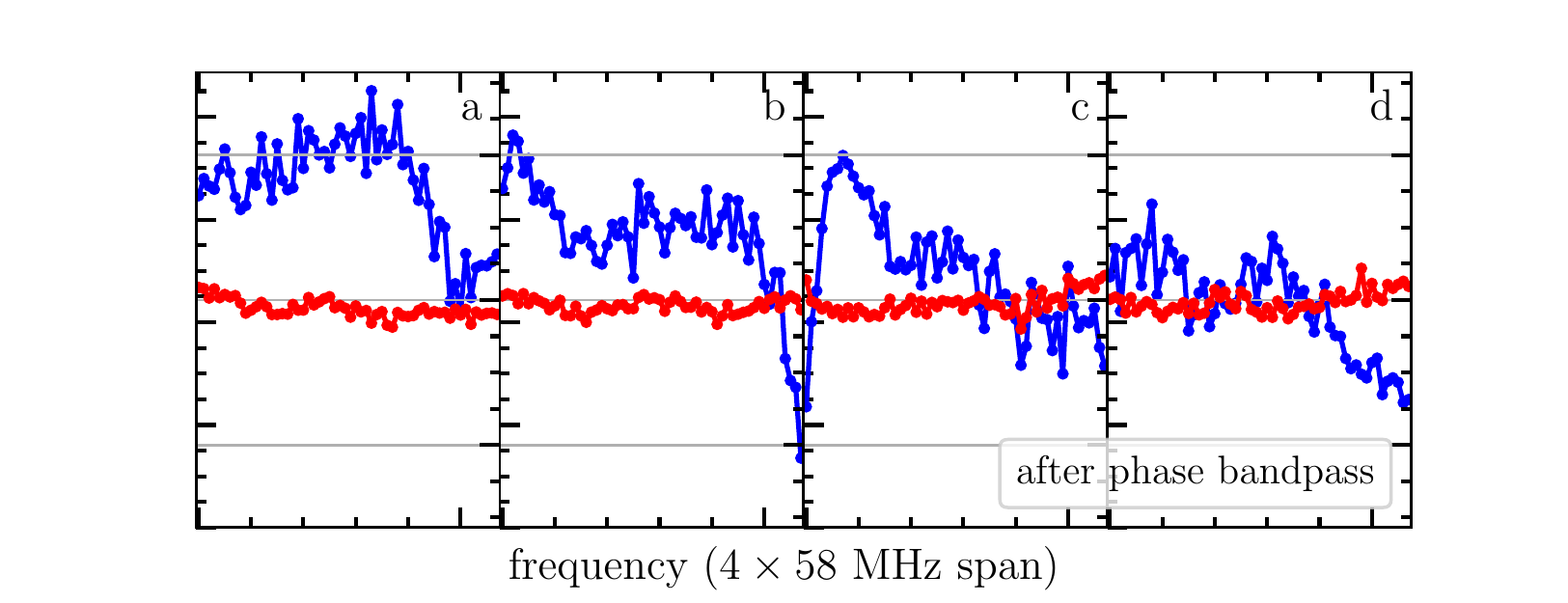}
    \caption{Amplitude (blue) and phase (red, covering $\pm\pi$) spectrum of the correlation coefficient between the
Fort Davis VLBA station (FD) and the Green Bank Telescope (GBT) during a scan on calibrator source 1749$+$096,
before and after phase bandpass correction. The spectrum is shown across the
four 58\,MHz channels (labeled a,b,c,d) that are defined at correlation. The phase bandpass
correction over the GMVA 256\,MHz bandpass is small, but the correction is more
pronounced near the edge of the much wider 2\,GHz wide EHT bands. Using HOPS
control codes, the pipeline is able to correct for an offset, as well as one
slope (single-band delay) per channel.}
\label{fig:spectrum}
\end{figure}

Because \texttt{fourfit} output is already channel averaged, it is not possible to directly measure {\em intra}channel phase bandpass from detected fringes, regardless of S/N. Generally the phase evolution across each 58\,MHz channel is small ($< 10^\circ$), as is any possible coherence loss from residual intrachannel phase variation. To track situations of more rapid intrachannel phase variation, particularly near the 2\,GHz band edge of the EHT, the first-order phase slope $\partial\phi_{0i}/\partial\nu$ is also estimated using the differences between nearby channels, and a linear phase slope correction is implemented as a channel-by-channel ``single-band-delay'' (SBD) offset referenced to the center of each channel.

The total instrumental phase attributed to each station $j$ relative to the reference station is
\begin{equation}
    \theta_j(f) = \phi_{0j,c} + 2\pi\,(f-f_{\mathrm{ref},c})\,\tau_{0j,c},
\end{equation}
where within the range of each channel $c$, $\phi_{0j,c}$ is the average measured instrumental phase for channel $c$ taken at the channel reference frequency $f_{\mathrm{ref},c}$, and $\tau_{0j,c}$ is a small single-band delay used to track phase variation within each 58\,MHz channel. Due to the available tuning parameters in \texttt{fourfit}, the $\phi_{0j,c}$ contribution is polarization dependent, while the $\tau_{0j,c}$ contribution is taken as an average over both polarizations. An example of phase (and amplitude) bandpass for a GMVA baseline between Fort Davis and GBT at 86\,GHz is given in \autoref{fig:spectrum}, before and after correction using the piecewise parameterization available in HOPS.

\subsection{Atmospheric Phase}
\label{sec:adhoc}

\begin{figure}
    \centering
    \includegraphics[width=1.0\linewidth]{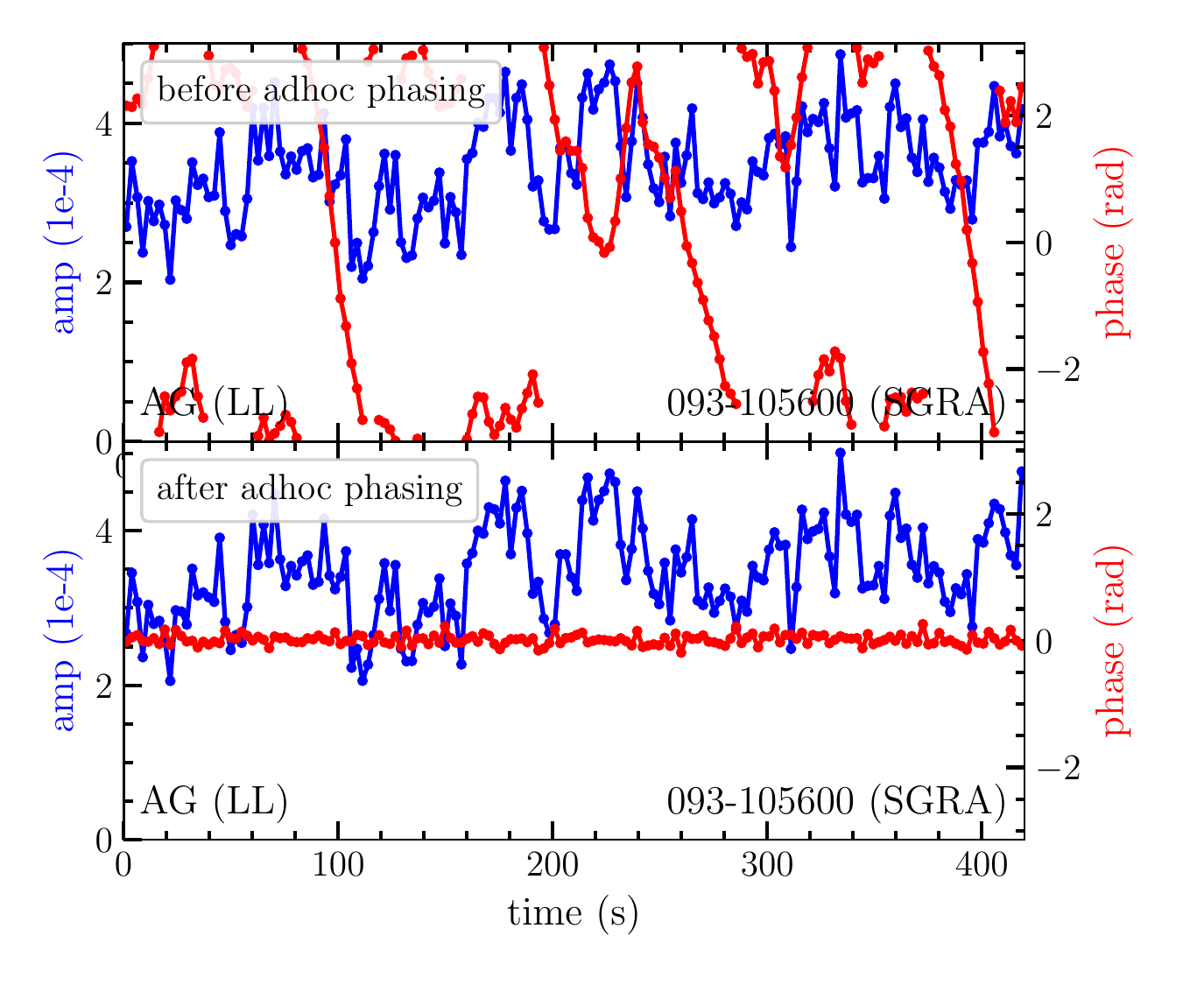}
    \caption{Amplitude (blue) and phase (red) time series of the correlation coefficient between
ALMA and GBT during a scan on \sgra. Atmospheric phase compensation is done
using the round-robin implementation from \autoref{sec:adhoc} that prevents
self-tuning, with an automatically chosen effective integration timescale $\calT{dof} = 2.5$\,s.}
\label{fig:timeseries_adhoc}
\end{figure}

The phase evolution over time captured by the first-order fringe fit is insufficient for millimeter VLBI, where atmospheric turbulence causes nonlinear, stochastic phase evolution on timescales of seconds (\autoref{fig:timeseries_adhoc}), much shorter than a typical VLBI scan length of minutes. Unlike the nonlinear corrections in phase from stable instrumental bandpass mentioned previously, atmospheric phase is continually changing and must be measured and corrected {\em on-source}. HOPS provides the ability to pre-correct nonlinear phase evolution over time using station-based ad hoc phases, where the term ad hoc is used to distinguish these arbitrary atmospheric phase corrections from the modeled linear phase drift due to delay rate. These nonlinear corrections are estimated and applied at stage~3 in the pipeline, resulting in an overall increase in scan-average S/N, as well as increased precision and overall self-consistency of the linear fringe solutions across the array.

Nonlinear time-dependent phase in the EHT-HOPS pipeline is estimated per scan using on-source detections from a single reference station to other stations in the array. The correlated signal must be strong enough so that phase can be estimated on a timescale that is short with respect to the atmospheric coherence time. Corrective phases relative to the reference station are then applied to the $N-1$ remaining stations, stabilizing relative phase due to station-based variation across the entire array. The following subsections describe how the reference station is chosen for each scan and how the data are stacked for short-timescale phase estimation. A stochastic atmospheric phase model of a known power spectrum is assumed so that the variation from atmospheric phase drift can be balanced against available S/N in the data. This sets the effective integration timescale for phase estimation, which is then performed in a round-robin estimation/application process to avoid self-tuning on statistical noise.

\Needspace*{4\baselineskip}
\subsubsection{Reference Station Selection}

Similar to instrumental phase, atmospheric phase corrections are assigned to each station relative to a reference station. However, since one reference antenna may not be present in all scans, the choice of reference antenna is made scan by scan by maximizing a statistic designed to capture the total measurable phase degrees of freedom using only baselines to the reference antenna. The scoring depends on the S/N from the proposed reference antenna to all remaining antennas $\rho_{0i}$, the S/N required for a good phase measurement $\rho_\mathrm{dof} \sim 10$, an S/N threshold below which false fringes appear $\rho_\mathrm{thr} \sim 7$, and an assumed phase coherence timescale $\calT{coh} \sim 6$\,s at 1.3\,mm and $\sim$18\,s at 3.5\,mm, characteristic of challenging weather. Here $\calT{coh}$ is defined as the expected time span over which phase drifts by 1\,rad (as defined later in \autoref{eqn:structfunc}).

Each $\rho_{0i}$ between a possible choice of reference station~$0$ and remaining station $i$ is taken as a quadrature sum of the individual $\rho_{0i,j}$ from each of four polarization products $j$, reflecting the fact that changes in atmospheric path delay do not depend on polarization and polarization products can be stacked for better S/N:
\begin{equation}
    \rho_{0i}^2 = \sum_j{\rho_{0i,j}^2 (2/\pi)\arctan\left[
    (\rho_{0i,j} / \rho_\mathrm{thr})^4\right]}.
\end{equation}
The arctan logistic function quickly transitions from 0~to~1 as $\rho_{0i,j} > \rho_\mathrm{thr}$, and is used to apply a threshold $\rho_\mathrm{thr}$ below which to ignore likely false fringes.
For a given baseline $0$--$i$ with scan-average S/N of $\rho_{0i}$,
we estimate the number of segments that could be formed by
splitting the scan in time, while maintaining an S/N above $\rho_\mathrm{dof}$ for each segment. This corresponds to the number of measurable degrees of freedom above some nominal statistical precision
\begin{equation}
    N_{\mathrm{meas},0i} = (\rho_{0i} / \rho_\mathrm{dof})^2.
\end{equation}
At very high S/N, the number of measurable degrees of freedom might be very large, corresponding to a very short segment duration.
In this situation, the maximum useful degrees of freedom over total duration $\mathcal{T}_{0i}$ are limited by the number of phase measurements required to fully characterize any atmospheric variability. Correcting phase more rapidly than $\sim$5 times per $\calT{coh}$ gives rapidly diminishing returns, so we set
\begin{equation}
    N_{\mathrm{max},0i} = 5 \,(\mathcal{T}_{0i} / \mathcal{T}_{\mathrm{coh},0i}).
\end{equation}
Finally, we calculate the total useful degrees of freedom by summing over all baselines to the proposed reference station under a scheme that reflects the diminishing returns for measurable degree of freedom beyond the maximum useful number
\begin{equation} 
    N_{\mathrm{useful},0} = \sum_i N_{\mathrm{max},0i} \ln(1 + N_{\mathrm{meas},0i} / N_{\mathrm{max},0i}).
\end{equation}
The reference station chosen for each scan and set as station $0$ is the one that provides the largest $N_{\mathrm{useful},0}$ for detections from that scan.

\subsubsection{Data Alignment}

When the time required to accumulate $\rho_\mathrm{dof}$ approaches $\calT{coh}$, performance of on-source phase stabilization increases dramatically by stacking data prior to measuring phase. For example, by stacking two equal-sensitivity measurements and increasing the S/N by $\sqrt{2}$, the timescale over which phases can be reliably estimated is correspondingly reduced by a factor of two. 

For the purpose of atmospheric phase estimation, the EHT-HOPS pipeline stacks data from all polarization products by aligning the data empirically before computing a weighted average. First, data are band averaged and adjusted to a common fringe solution, as prior to step~5 fringe globalization (\autoref{fig:stages}), different polarization products may have different delay rate solutions. The empirical phase offset between one of the polarization products $r_i$ and another $r_j$ is measured by segmented average
\begin{equation}
    \Delta\phi_{ij} = \mathrm{Arg}\left[\sum_n {r_i[n]\, r_j^\ast[n]}\right],
\end{equation}
where the length of each segment should be long enough to accumulate to $\rho > 1$.

The measured $\Delta\phi$ between one of the polarization products and others is then used to align and stack the original visibility data. While it may be challenging to accumulate sufficient S/N per $\calT{coh}$, the S/N across an entire scan is many times larger so that $\Delta\phi$ is accurately measured. The same alignment procedure can be used to stack data from multiple independently processed bands when available.

\subsubsection{Phase Model}

Atmospheric phase is assumed to follow a random stochastic process due to a turbulent cascade. In this section we adopt a phase model appropriate for a single station, although the atmospheric phase corrections will cover the combined effects on a baseline. The model itself is used to set tuning parameters and needs to reflect broadly the ensemble behavior rather than be exact. The phase variation is captured by the phase structure function, which typically follows a power-law profile over a wide range of scales
\begin{equation}
    D_\phi(t) = \E{[\phi(t'+t)-\phi(t')]^2} \approx (t/\calT{coh})^\alpha.
    \label{eqn:structfunc}
\end{equation}
In this representation, the coherence timescale $\calT{coh}$ is the time after which phase is expected to drift on average by 1\,rad. The power-law index $\alpha$ will be modified at large scales (where energy is injected) and small scales (where energy is dissipated), but these limits are typically outside the primary timescale range of interest -- from the minimum useful integration time up to the duration of a scan. For 2D Kolmogorov turbulence $\alpha=2/3$, and for 3D Kolmogorov turbulence $\alpha=5/3$ \citep{tms}. Measured values of $\alpha$ generally lie somewhere in between. The corresponding power spectrum is
\begin{equation}
    S_\phi(f) = \Gamma[1+\alpha] \sin\left(\frac{\pi\alpha}{2}\right) \calT{coh}\, (2\pi \calT{coh} f)^{-1-\alpha},
\end{equation}
which is related to the structure function through the autocorrelation function
\begin{equation}
    C_\phi(t) = \E{[\phi(t'+t)\phi(t')]^2} = C_\phi(0)-D_\phi(t)/2, 
\end{equation}
and its Fourier transform
\begin{equation}
    S_\phi(f) = \int_{-\infty}^{\infty}dt\, C_\phi(t)\,e^{-2\pi i ft}.
\end{equation}

Phase estimation is done using an atmospheric phase model drawn from a Savitzky--Golay (\texttt{savgol}) filter \citep{savitzky1964smoothing} applied to the visibility data, which is a running piecewise polynomial fit that has a convenient implementation in Scipy. The filter acts as a symmetric low-pass linear filter for regularly spaced data \citep{schafer2011savitzky}. Real and imaginary components of the complex visibility time series are filtered separately, and the filtered visibilities are used to derive a smoothly varying interpolated phase estimate over time at the location of each data value.

The \texttt{savgol} filter fits an $n$-degree polynomial over a window length $\calT{win}$, so that the effective integration time $\calT{dof}$ per degree of freedom is $\calT{win}/(n+1)$. 
The statistical phase noise for a measurement taken over $\calT{dof}$ is approximately
\begin{equation}
    \sigma_{\phi,\mathrm{thermal}}^2 \approx (\rho_1^2 \calT{dof})^{-1},
\end{equation}
where $\rho_1$ is the S/N in 1\,s of accumulation, and we have ignored impact on S/N from coherence loss from atmospheric phase drift over the integration period.

In addition to statistical noise, there is residual phase noise from the inability of the smoothed model to capture true rapid phase variations. The residual noise after filtering by window function $w(f)$ can be calculated from integrating residual power in the frequency domain:
\begin{equation}
    \sigma_{\phi,\mathrm{residual}}^2 = \int_{-\infty}^{\infty}df\, [1-w(f)]^2 \, S_\phi(f).
\end{equation}
For a boxcar moving-average filter of length $\calT{dof}$ (equivalent to \texttt{savgol} filter of degree zero), the window function is $\mathrm{sinc}(\pi f \calT{dof})$ and the residual power is
\begin{equation}
    \sigma_{\phi,\mathrm{residual}}^2 = 
    \frac{2^{-\alpha}(2+\alpha-2^\alpha)}{(1+\alpha)(2+\alpha)}
    \left(\frac{\calT{dof}}{\calT{coh}}\right)^\alpha.
\end{equation}
Other window functions such as ideal low-pass and Gaussian give equally simple expressions, and all scale as $(\calT{dof}/\calT{coh})^\alpha$. The boxcar response is a reasonable approximation to \texttt{savgol} filters of low nonzero degree.

The effective averaging time $\calT{dof}$ that minimizes total error $\sigma_{\theta,\mathrm{thermal}}^2 + \sigma_{\theta,\mathrm{residual}}^2$ is
\begin{equation}
    \calT{dof} = \left[\frac{(1+\alpha)(2+\alpha)}
    {2^{-\alpha}\alpha(2+\alpha-2^\alpha)\rho_1^2}
    \calT{coh}^\alpha\right]^{1/(\alpha+1)},
\end{equation}
which is close to the $\calT{dof}$ where $\sigma_{\phi,\mathrm{thermal}}^2 = \sigma_{\phi,\mathrm{residual}}^2$. We use this optimal $\calT{dof}$ to set the parameters of the \texttt{savgol} filter within the constraints of the filter construction (filter length $N_\mathrm{savgol}$ in units of the correlator accumulation period $\calT{AP}$ is odd and equal to or greater than polynomial degree $d$),
\begin{equation}
    N_\mathrm{savgol} = \max\left\{1+d,
    1+2\,\lfloor\frac{(1+d)\,\calT{dof}}{2\, \calT{AP}}\rfloor\right\} .
\end{equation}

\subsubsection{Round-robin Implementation}
\label{sec:roundrobin}

Atmospheric phase compensation requires a large number of parameters to be derived from data on-source in a regime that is often S/N limited. By restricting the number of fitted degrees of freedom based on available S/N, the previously outlined strategy helps to avoid introducing additional noise from over-fitting to mere statistical variations. However, some degree of fitting to thermal noise is inevitable, and this can lead to biases in derived quantities -- such as a positive bias in coherently averaged visibility amplitude through the introduction of false coherence. To avoid biases from self-tuning, \citet{Johnson2015} estimate from and apply phases to data corresponding to different polarization products.

We employ a round-robin (leave-out-one) scheme for phase corrections that partitions over frequency to ensure that any phase adjustments are derived from data that are disjoint from the data they are being applied to. Because path variations due to the atmosphere are expected to be achromatic over the observing bandwidth, visibilities for each of the $N$ frequency channels can be phase stabilized using a smooth atmospheric phase model derived from the remaining $N-1$ channels. As long as the number of channels in the data is large (EHT bands are partitioned into 32 corresponding ALMA channels for correlation), the leave-out-one strategy uses most of the available S/N for estimating a phase model and avoids entirely issues of self-tuning. One drawback to the strategy is that it does not transition naturally to making one stable common phase adjustment to all channels in the limit of low S/N (or no correction at all), which is the desired behavior. Atmospheric phase correction at 86\,GHz is demonstrated in \autoref{fig:timeseries_adhoc}, where a strong baseline between ALMA and GBT is able to self-correct phases at a timescale of 2.5\,s while using the round-robin approach over four independent channels.

\Needspace*{4\baselineskip}
\subsubsection{Second-order Corrections}

Because it is the atmospheric path length variations and not phase variations that we assume are achromatic, a small frequency-dependent adjustment is made to the original unwrapped phase corrections based on the relative difference of the channel frequency to a reference frequency (typically set to the middle of the entire band, and assumed to be representative of the frequency at which estimates are made)
\begin{equation}
    \phi \to \phi\,\frac{f_\mathrm{chan}}{f_\mathrm{ref}} .
\end{equation}
The adjustment can be interpreted as tracking the small nonlinear variations in delay that are inferred from measured phase drift.

Residual frequency offsets in the data can also be corrected at this stage through explicit frequency shifting, so long as the frequency shift $\delta f$ is small compared to the sampling of the data,
\begin{equation}
    \phi \to \phi + 2\pi\delta f\,t .
\end{equation}
If left uncorrected, the effects of the frequency offset will instead be fit through a delay rate compensation $\delta\dot{\tau}$, through the association $\delta f \leftrightarrow \nu \delta\dot{\tau}$. However, since the residual fringe rate $\nu \delta\dot{\tau}$ varies with observing frequency while the frequency offset $\delta f$ is fixed, the corrections are not identical and the compensation through fringe rate (essentially stretching or compressing the data in time) imprints a second-order effect that scales with the fractional bandwidth. Thus, it is best to measure any residual frequency offset and correct it at correlation or in data pre-processing prior to fitting delay rate.

\subsubsection{Comparison to standard techniques}

Stochastic phase variation due to atmospheric turbulence is a dominant residual systematic for high-frequency VLBI observations, and the success of on-source phase estimation and compensation is a major factor in the quality of fringe fitting and reduction. Traditionally, this is handled by dividing data into segments shorter than the phase coherence timescale. The complex correlation coefficients can then be vector averaged for each individual segment without suffering much decoherence from drifting phase.

Baseline measurements for each segment can be used to reference phases to a single antenna under a global fringe solution that includes absolute station phase \citep[e.g., ][]{schwab1983}, or they can be used to form derivative products such as closure phase \citep{rogers1974} and closure amplitude \citep{Readhead1983} that cancel out station gains and are sensitive to only source structure. Phase referencing to a reference station will try to transfer any unmodeled structure phase to baselines that do not include the reference antenna. In this way, one expects similar results if forming a closure phase from multisegment averages of phase-corrected visibilities, or if averaging many closure phases that are themselves calculated individually for each segment, aside from details related to nonlinear propagation of thermal noise at low S/N \citep{Rogers1995}.

For the on-source atmospheric phase calibration presented here, we have incorporated 1) automated selection of reference station based on available S/N across the array; 2) coherent stacking of polarization products for increased S/N during phase estimation; 3) corrective phases which are estimated smoothly over the scan, using an adaptive effective integration time that balances statistical errors to those from expected residual phase drift; and 4) a strategy to avoid self-tuning on statistical fluctuations while still using most of the data for estimation. Alternate strategies for S/N-dependent selection of integration time \citep{janssen2019} and cross-application of estimated phases \citep{Johnson2015} have been presented elsewhere. The use of a \texttt{savgol} filter for smooth estimation of local complex visibility prior to phase estimation is similar to the use of overlapping segments by \citet{Rogers1995} in the context of incoherent averaging of amplitude. The standard approach of using independent segments of vector-averaged visibilities can be considered a down-sampled version of a boxcar moving-average coherent integration window. The boxcar window (\texttt{savgol} order zero) acts as a low-pass filter with a sinc response, while higher-order \texttt{savgol} filters will have a sharper cutoff.

\subsection{RCP--LCP Delay Calibration}
\label{sec:rldelay}

Signals that take different analog paths from the receiver to recording elements will be subject to different delays from cables, clocks, and electronics. The sensitivity of the measured correlation to relative delay depends on the inverse bandwidth -- for 1\,GHz of bandwidth, the relative delay should be known to much better than 1\,ns for sufficient coherence across the band. It is particularly important to delay align RCP and LCP feeds at each antenna, to be able to stationize the (polarization-independent) atmospheric and geometric delay across all four polarization products. It can also be useful to estimate stable instrumental relative delays between frequency bands so that a station-based set of delays is characterized by only a single free delay parameter per station instead of one per station per band. Because components of the receiving system and electronics are generally locked to the same clock reference, instrumental contributions to delay rate are generally not polarization dependent and do not need to be relatively calibrated. For the same reason, the instrumental delay calibration is generally stable over time so long as the setup is not disturbed.

During the initial stages of the EHT-HOPS pipeline, the fringe search is unconstrained within some delay (and delay rate) search window that is wide enough to accommodate the full range of residual geometric, atmospheric, instrumental, and clock errors. Each baseline and polarization product is fit separately to a relative delay and delay rate. One strategy to align R--L delay at an antenna $j$ is to measure the relative delay to RCP and LCP feeds at a site given a common reference signal (e.g. LCP at some other station $i$). This requires some amount of linear polarization in the source to produce a cross-hand fringe in addition to the parallel-hand fringe. Then, for example,
\begin{equation}
    \tau_{j,\text{R--L}} = \tau_{ij,\mathrm{LR}} - \tau_{ij,\mathrm{LL}}
\end{equation}
with $\tau_{ij,\mathrm{LR}}$ and $\tau_{ij,\mathrm{LL}}$ as the measured baseline relative delays measured for polarization products LR and LL, and $\tau_{j,\text{R--L}}$ the inferred relative delay between RCP and LCP at station $j$. The measurement can be averaged over all available reference signals for increased accuracy.

One drawback to the reference signal strategy is that detected fringes in the cross-hand polarization products are sensitive to polarization leakage since both the typical magnitude of leaked power and the degree of linear polarization are often of the same magnitude. Therefore, prior to polarization leakage calibration, parallel-hand correlated signal can leak into the cross-hand measurement and introduce significant noise in the delay measurement.

When ALMA is present in the array, it can be used to measure RCP--LCP delay at other stations using only parallel-hand products to ALMA. This is because the ALMA linear feeds are delay and phase aligned through ALMA quality assurance calibration \citep{Goddi_2019}. The PolConvert process converts ALMA's mixed-polarization products to circular polarization, maintaining the zero relative delay between ALMA-converted RCP and LCP \citep{Marti_2016}. Then,
\begin{equation}
    \tau_{j,\text{R--L}} = \tau_{\mathrm{A}j,\mathrm{RR}} - \tau_{\mathrm{A}j,\mathrm{LL}}
\end{equation}
with A for ALMA. Since the R--L instrumental delays are generally stable through the night, ALMA only needs to be present in a subset of scans in order to fully \mbox{R--L} delay calibrate the network. The basic strategy at stage~4 of the pipeline is therefore to take an average of ALMA parallel-hand detections to other stations to derive a single RCP--LCP delay offset for each non-ALMA site on each observing night. The average itself is a $1/\sigma^2$ weighted mean, after accounting for a small amount of systematic delay error and after rejecting 10\,$\sigma$ outliers from the median value. Further validation steps check that the constant offset is a good model to within thermal error plus small systematic tolerances.

\begin{figure}
    \centering
    \includegraphics[width=0.95\linewidth]{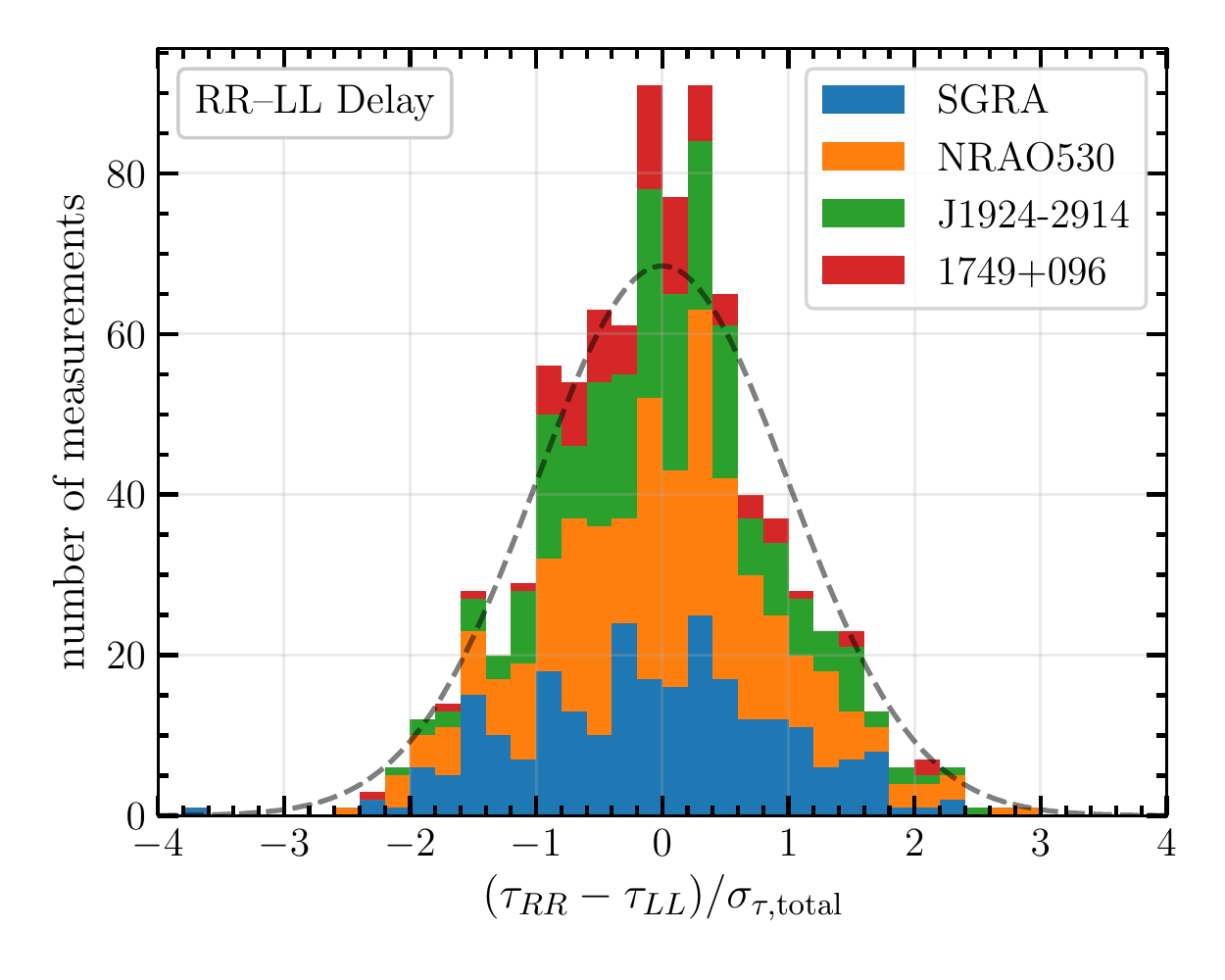} \\
    \includegraphics[width=0.95\linewidth]{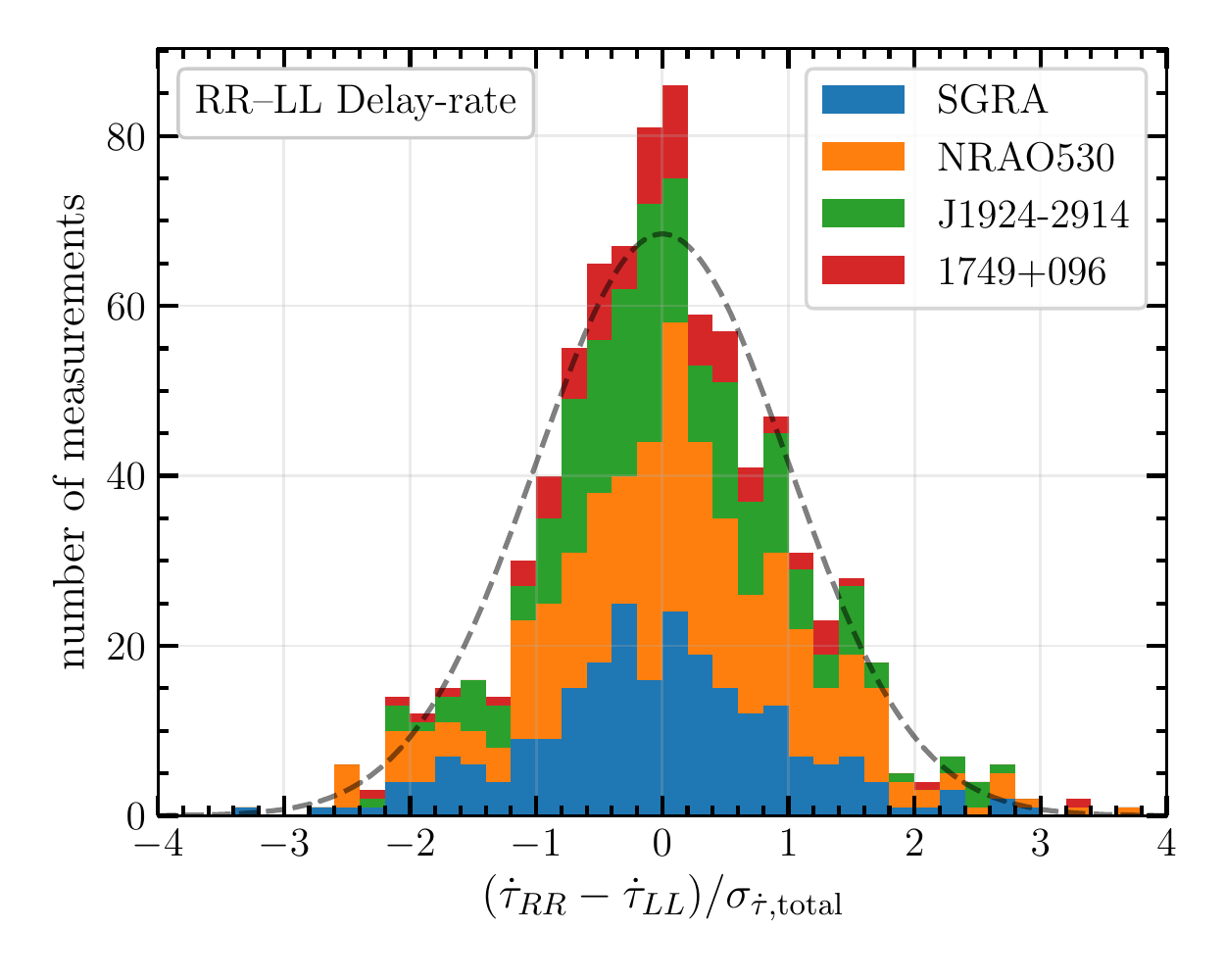}
    \caption{Distribution of differences between calibrated RR and LL measured delays and delay rates for
all scans in the test data set with S/N $>$ 7, in units of expected total measurement error for the difference (taken in quadrature) from \autoref{eqn:delayerr} with 10 ps
systematic in delay and 0.25\,fs/s in delay rate. The dotted line corresponds to a standard normal distribution, expected if the constant delay model is valid to within total Gaussian measurement uncertainties. 10\,ps corresponds to a negligible $4\times10^{-5}$ fractional coherence loss over the 256\,MHz GMVA bandwidth, and the same for a residual 0.25\,fs/s delay rate error over 240\,s of integration at 86\,GHz (\autoref{eqn:fractionalloss}). The successful fitting of parallel-hand delay differences using a constant model
with zero statistically significant outliers indicates good parallel-hand fringe
solutions for both strong and weak sources, as well as the stability of
individual station instrumental delays through the night. There are a small number (2\%) of delay rate difference
outliers owing to the current limitation of ad hoc phasing to a single reference station, meaning that some weaker isolated
baselines may not be able to be phase stabilized. A single station delay rate is still enforced at the global fringe
solution, but for the original delay rate outliers there could be residual coherence loss under
a full scan average.}
\label{fig:rrll_distribution}
\end{figure}

\autoref{fig:rrll_distribution} shows the distribution of all measured \mbox{RR--LL} delay differences between ALMA and other stations after calibrating out a constant delay offset between RCP and LCP feeds at non-ALMA sites. The fact that all measured differences are consistent with zero confirms the assumed stability of RCP versus LCP relative delay at each site.

\subsection{Global Fringe Solution}
\label{sec:global}

After stage~4, fitted delays and delay rates on each baseline are expected to be the same for all polarization products to within measurement thermal noise. This allows us to {\em stationize} the fitted delay and delay rate parameters, modeling each as the difference between a pair of station delays and delay rates. The stationization of the fringe solution provides several benefits: it prevents the first-order fringe correction from introducing nonclosing (not station-based) phase adjustments to the data, it reduces the total number of free parameters describing the corrections from a number that scales with the $\mathcal{O}(N^2)$ baselines to a number that goes as $\mathcal{O}(N)$ stations, and it allows fringe locations to be accurately predicted on baselines that may have no independently detectable correlated signal.

The thermal contribution to errors in the estimation of delay and delay rate is directly related to the noise in a measurement of total accumulated phase drift across bandwidth $\Delta\nu$ and time $\Delta t$.
At moderate S/N and near the true fringe peak, the error is approximately $\sqrt{12}/\rho$, where $1/\sqrt{12}$ is the
standard deviation of a uniform distribution corresponding to the flat
integration period and $1/\rho$ represents thermal noise in the phase
measurement \citep[A12.25]{tms}. Therefore,
\begin{equation}
\sigma_{\tau,\text{thermal}} = \frac{1}{2\pi}\frac{\sqrt{12}}{\rho \, \Delta\nu} \quad
\sigma_{\dot{\tau},\text{thermal}} = \frac{1}{2\pi\nu}\frac{\sqrt{12}}{\rho \, \Delta t}.
\end{equation}
In addition to the thermal error, we can add some level of systematic error to fitted
delay and delay rate,
\begin{gather}
\label{eqn:delayerr}
\sigma_\tau^2 = \sigma_{\tau,\text{thermal}}^2 + \sigma_{\tau,\text{sys}}^2 \\
\sigma_{\dot{\tau}}^2 = \sigma_{\dot{\tau},\text{thermal}}^2 + \sigma_{\dot{\tau},\text{sys}}^2,
\end{gather}
which may be baseline and polarization dependent. The systematic errors arise
from search resolution and interpolation accuracy, contamination from leaked
signal power (particularly in cross-hand products), or other baseline-dependent
processing artifacts and in general must be estimated from the data.

\begin{figure*}[t]
    \centering
    \includegraphics[width=1.0\linewidth]{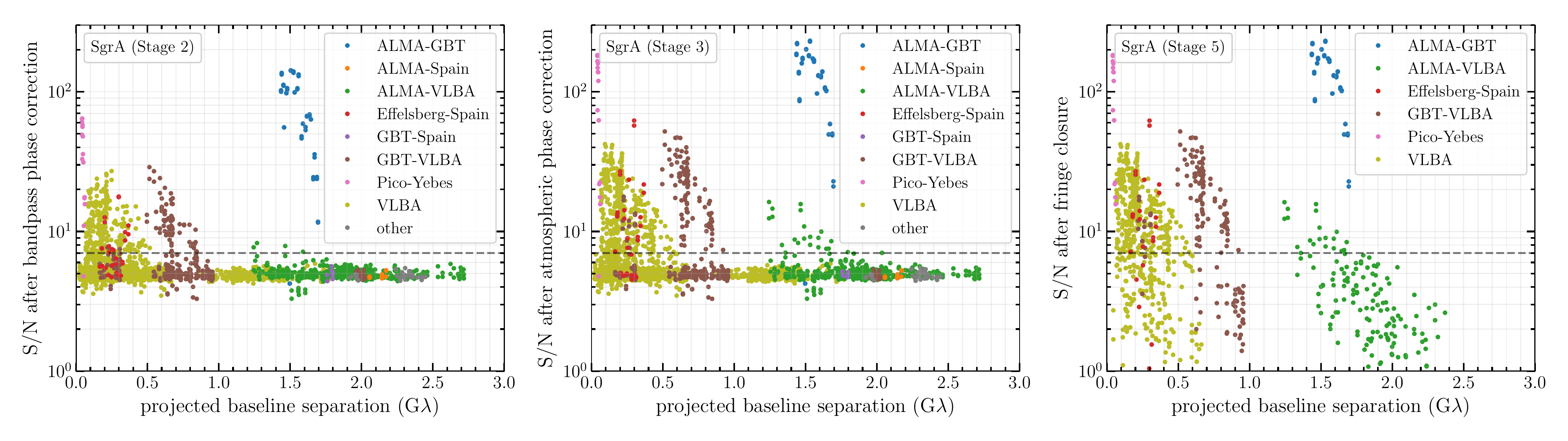}
    \caption{Scan-average S/N as a function of projected $(u,v)$ distance for GMVA+ALMA 3.5\,mm observations of \sgra. The three panels represent fringe solutions from successive stages of the EHT-HOPS pipeline, before and after atmospheric phase correction and after a station-based global fringe solution has been applied. From the left to middle panel, the removal of nonlinear phase variations increases scan-average S/N by preserving phase coherence across the scan. Both stages, however, suffer from a characteristic false fringe S/N $\sim$ 5 from random noise fluctuations due to the large number of trials over the delay and delay rate search window. The horizontal dashed line at S/N = 7 represents a threshold at which fringe solutions are considered reliable (very unlikely to have arisen from a noise fluctuation). These confident detections are used to solve a global fringe solution across the entire array, reducing the number of effective trials for weak baselines to 1 when the fringe solution can be constrained by other detections. This allows measurement down to an uncertainty of S/N $\sim$ 1 (right panel), where we can see a clear decrease in S/N with increasing baseline length on long ALMA--VLBA baselines (generally connected through ALMA$\leftrightarrow$GBT$\leftrightarrow$VLBA).}
    \label{fig:3mm_sgra_snr}
\end{figure*}

For a fringe search that fits delay and delay rate to values which maximize
total detected fringe power $\rho_0^2$, we must also consider the probability
of false positive, i.e., one minus the probability that all of $N$ independent noise
measurements across the search space in delay and delay rate are less than the
measured value. Thermal noise gives an exponentially distributed
random contribution to fringe power, so the probability of a false positive
over $N$ trials \citep{tms} is
\begin{align}
\label{eqn:falsepositive}
\prod_i^N P(\rho_i < \rho_0) &= 1 - \left[1-\exp\left(-\frac{\rho_0^2}{2}\right)\right]^N \\
    &\approx N\exp\left(-\frac{\rho_0^2}{2}\right).
\end{align}
This is also the cumulative (survival) distribution of the maximum noise fringe
power over $N$ independent measurements. A requirement that the false-positive
rate be very low (much less than the number of fits performed) sets a threshold
$\rho_\text{thr} \sim 7$ above which detections are considered reliable.

Following \cite{alef1986}, we take the estimated baseline and
polarization-dependent delay and delay rate solutions along with their errors from
\autoref{eqn:delayerr}, and then perform a least-squares fit to station-based
parameters. For each scan, one delay and delay rate parameter is fit per
antenna that minimizes the squared error across all baseline measurements.
Measurements with $\rho_0 < \rho_\text{thr}$ are assigned a very large
$\sigma_\text{sys}$ so that they are effectively ignored in the presence of any
other constraining data. The least-squares minimization is performed in
Scipy with an additional
\texttt{soft\_l1} loss function $L(z^2) = 2f^2\,(\sqrt{1 + z^2/f^2} - 1)$ applied at scale $f = 8\,\sigma$ to mitigate the effects of
outliers.

Specifically the least-squares approach solves for, e.g., model station delays $\tau_i$ (and delay rates $\dot{\tau}_i$) by minimizing a chi-square error function
\begin{equation}
    \chi^2 = \sum_{i<j,\,k} L\left[\frac{(\tau_{ij,k} - (\tau_{j} - \tau_{i}))^2}
    {\sigma_{\tau,ij,k}^2} \right]
\end{equation}
where $i<j$ loops over baselines, $k$ indexes the four polarization products, $\tau_{ij,k}$ is the measured delay for each baseline/polarization, $\tau_i$ and $\tau_j$ are not dependent on polarization owing to the previous step of delay calibration, $\sigma_\tau$ is total error as described in \autoref{eqn:delayerr}, and $L(z^2)$ is the \texttt{soft\_l1} loss function specified earlier, as implemented in Scipy.
The best-fit station delays and delay rates are used to model baseline fringe parameters
\begin{equation}
\tau_{ij} = \tau_j - \tau_i \quad \text{and} \quad \dot{\tau}_{ij} = \dot{\tau}_{j} - \dot{\tau}_{i},
\end{equation}
which are then applied to the data for the global fringe solution (as zero-width
search windows).

Expanding the fringe amplitude (\autoref{eqn:fringe}) to second order
about zero total phase drift,
\begin{equation}
\frac{|r_\text{off-fringe}|}{|r_\mathrm{ideal}|} \sim 1 - \frac{(\Delta\phi)^2}{24}
\end{equation}
so that a total phase drift of $\sqrt{0.24}$\,rad corresponds to a 1\%
amplitude loss. The expected amplitude loss at a fringe solution based on a
measurement of S/N $\rho$ is $1/(2\rho^2)$ each for delay and rate
errors and not including any noise bias. Propagating fringe solutions with
S/N of 7 and above will maintain sub-1\% amplitude loss.

In terms of errors on delay $\tau$ and rate $\dot\tau$ directly, the amplitude
efficiency loss factor is
\begin{equation}
\frac{(2\pi\tau\Delta\nu)^2}{24} \qquad \frac{(2\pi\nu\dot\tau\Delta t)^2}{24}.
\label{eqn:fractionalloss}
\end{equation}
To maintain sub-1\% amplitude loss for an observing bandwidth of $\Delta\nu =
2$\,GHz at observing frequency $\nu = 220$\,GHz, delay must be within 0.04\,ns and
rate must be within 0.07\,ps/s for $\Delta t = 10$\,s coherent integration.
These limits are within typical systematic errors seen in real data (e.g.,
\autoref{fig:rrll_distribution}).

Not all baselines are constrained by the global fringe solution. If a station
has no reliable ($\rho_0 > \rho_\text{thr}$) detections to other stations in
the array, its relative delay and delay rate to other sites remain unconstrained.
Following fringe globalization, stations in the array are partitioned into fringe groups. Each group represents a
set of mutually connected stations, where stations are connected through one or more
baselines where at least one polarization product gives fringe detection
with $\rho_0 > \rho_\text{thr}$. Baselines between stations that belong to different
fringe groups are flagged from the data and removed, so that the only surviving correlation
measurements are those that are evaluated at single well constrained fringe locations.
After the global fringe solution is adopted, individual baselines that have well constrained fringe parameters can be measured to arbitrarily low S/N, as shown in \autoref{fig:3mm_sgra_snr}, and are no longer subject to a noise floor owing to the large fringe search parameter space.

As noted by \citet{alef1986}, the least-squares global fringe fit derived from initial
baseline-based fringe solutions is not as powerful (in terms of optimal S/N) as the coherent global fringe search of \citet{schwab1983}. However, for our purposes it offers a few advantages. For one, the baseline-based fringe search with independent solutions per baseline and per polarization product is extremely useful for using delay consistency to test for instrumental artifacts, data issues, and false fringes, for which \autoref{fig:rrll_distribution} provides one example. Second, the baseline-based fringe solution is immune to biases toward an assumed source model \citep{Wielgus2019}, as is does not use a source model. We note that the round-robin strategy as outlined in \autoref{sec:roundrobin} could also be used to avoid amplitude and phase biases in the Schwab-Cotton method. The difference in sensitivity between the baseline-based search and coherent global fringe search is not large for the EHT and GMVA because both arrays have relatively few stations (the difference between the $\mathcal{O}(N^2)$ baseline fit parameters and $\mathcal{O}(N)$ station parameters is not so large and can be made up by other optimizations such as optimal fringe solution intervals) and because both arrays are highly heterogeneous, with fringe solutions driven primarily by baselines to anchor stations such as ALMA or GBT.

\Needspace*{4\baselineskip}
\section{Post-processing}
\label{sec:postproc}

The EHT-HOPS pipeline is naturally divided into the initial stages 1--5, where iterations of HOPS \texttt{fourfit} fringe fitter are performed with increasing refinement of the initial phase calibration and a series of post-processing stages 6--9 that operate on the \texttt{fourfit} output. The first step (stage 6) in post-processing is to convert the \texttt{fourfit} native binary output data into standard \texttt{UVFITS} \citep{uvfits} format, using interfaces developed as part of the \texttt{eat} library for accessing and interpreting the HOPS Mark4 file set, as well as \texttt{UVFITS} interfaces originally developed for use in the \texttt{eht-imaging} library \citep{chael2016,Chael2018}. The data conversion routines are packaged as part of the \texttt{eat} library and provide a direct conversion of the HOPS ``type-2 fringe'' files into corresponding 
\texttt{UVFITS} format. The fringe files include all calibration corrections from stages 1--5 of the EHT-HOPS pipeline applied, including the fringe solution and atmospheric phase corrections, and are provided at the channel-averaged ``fourfit output'' time and frequency resolution described in \autoref{fig:accumulation}. This level of averaging is maintained until the final network calibration stage 9, at which data are further averaged (typically full-band, 10\,s averages) for a more convenient data volume. The post-processing stages 7--9 that follow read and write \texttt{UVFITS} formatted data directly and apply amplitude calibration and polarization- and time-dependent phase corrections that currently cannot be applied upstream within the HOPS framework. Because they operate on standard \texttt{UVFITS} output, the post-processing routines have some general utility even outside the EHT-HOPS pipeline.

\subsection{A Priori Flux Density and Field Angle Calibration}
\label{sec:apriori}

The sensitivity of each telescope is expressed by the SEFD, which represents the flux (Jy) of an unpolarized astronomical source that would be necessary to produce a received power equal to the system noise power (as in \autoref{eqn:sefdbp}). It can be estimated from observations of bright primary calibration targets (such as planets) and a calibrated measurement of atmospheric and receiver noise. At millimeter wavelengths, atmospheric noise and attenuation due to opacity are often substantial, so that SEFD can have a strong dependence on elevation. 

The EHT-HOPS pipeline relies on SEFD information delivered from each telescope, provided in the form of ANTAB\footnote{\url{http://www.aips.nrao.edu/cgi-bin/ZXHLP2.PL?ANTAB}} formatted data tables. The ANTAB tables provide SEFD information in the form of a constant degrees-per-flux-unit (DPFU) value encoding dish area and efficiency, an elevation-dependent parameterized gain curve efficiency correction $\eta_\mathrm{el}$, and a time-dependent effective system noise temperature $\T{sys}^\ast$ scaled according to the expected level of atmospheric attenuation through line-of-sight opacity $\tau$. While millimeter observatories generally estimate $\T{sys}^\ast$ directly via the ``hot-load" calibration technique \citep{Penzias_1973,Ulich_1976}, centimeter-wave observatories, such as the majority of stations in the GMVA \citep[even while observing at $\sim$several millimeters, see][]{martividal2012}, measure the system noise temperature $\T{sys}$ directly from calibrating to a noise diode injection, which does not account for atmospheric attenuation. In this case, opacity $\tau$ is estimated in the line of sight using the measured $\T{sys}$ and an estimate of the receiver noise temperature $\T{rx}$ and physical temperature of the atmosphere $\T{atm}$ \citep[e.g.,][]{Altshuler_1968}
\begin{equation}
     \T{sys} \approx \T{rx} + (1-e^{-\tau})\,\T{atm},
\end{equation}
and then used to scale $\T{sys}^\ast = e^\tau\,\T{sys}$ appropriately, as described in \citet{Issaoun2017a}.

For each source, the $\T{sys}^\ast$ values are interpolated into the observation times, and SEFD is calculated as
\begin{align}
\mathrm{SEFD} = \frac{\T{sys}^*}{\mathrm{DPFU} \times \eta_\mathrm{el}}.
\end{align}
The SEFD calibration tables are then used to amplitude calibrate visibilities in the \texttt{UVFITS} formatted data to a physical flux scale within the \texttt{eat} post-processing framework,
\begin{equation}
    V_{ij} = \eta_Q^{-1}\,\sqrt{\mathrm{SEFD}_i\times \mathrm{SEFD}_j}\,r_{ij},
    \label{eqn:vij}
\end{equation}
where $r_{ij}$ is the correlation coefficient as in \autoref{eqn:rij}.

Apart from flux scaling of visibility amplitudes, this stage of calibration also corrects for the a priori polarimetric field rotation angle, i.e., the relative orientation of the feed with respect to a fixed direction on the sky. The effect manifests as a nonlinear, source- and time-dependent phase offset between RCP and LCP components at each station; see \autoref{fig:polcal}, top panel. The field rotation angle $\vartheta_j$ is generally a combination of the source parallactic angle at the location of the $j$th station and a possible contribution from elevation-dependent rotation due to the receiver mount type. The correction takes the form of a station-based polarization-dependent phase correction to \autoref{eqn:vij} to align RCP and LCP to a fixed orientation on the sky. The middle panel of \autoref{fig:polcal} shows the R--L phase offsets after applying the field rotation correction.

\subsection{RCP/LCP Polarimetric Gain Ratios Calibration}

In order to form total intensity Stokes $I$ visibilities, it is necessary to calibrate the phase and amplitude mismatch between the measured LCP and RCP components. For small Stokes $V$ component and small leakage coefficients, the LCP and RCP visibilities are approximately related as \citep[e.g.,][]{Roberts_1994}
\begin{equation}
V_{jk, \rm RR} \approx \left(\frac{g_{j,\rm R}}{g_{j,\rm L}} \right)\left(\frac{g_{k,\rm R}}{g_{k,\rm L}} \right)^\ast e^{-2i (\vartheta_j - \vartheta_k)}V_{jk, \rm LL} .
\label{eqn:polgains}
\end{equation}
The field rotation component $ \exp \left[ -2i (\vartheta_j - \vartheta_k)\right]$ is removed at the a priori calibration stage, \autoref{sec:apriori}, leaving the complex gain ratios $g_{\rm{R/L}} \equiv g_{\rm R}/g_{\rm L}$ to be accounted for.
Polarimetric gain ratio phases are particularly important to calibrate.
The RCP versus LCP phase stability is analogous to the RCP versus LCP delay stability (\autoref{sec:rldelay}), but more sensitive by a factor corresponding to the inverse fractional bandwidth (e.g., two orders of magnitude more sensitive). Thus, while relative RCP versus LCP instrumental delays are generally constant throughout the night, relative instrumental phases can exhibit some residual drift.

The station-based phase offsets, $\theta_{\rm R-L}(t) \equiv \Arg{g_{\rm R/L}}$, are modeled as polynomial functions of time and are estimated directly from \autoref{eqn:polgains} with respect to a~reference station.
If the reference station gain ratio $g_{\rm R/L}$ is known, or can be derived a priori, as is the case for ALMA, this enables absolute calibration of the electric vector position angle on the sky for the entire array.
While the polynomial fit parameters are estimated from the data using robust, S/N-weighted statistics, the algorithm requires a manual selection of a polynomial degree used for a phase offset fit $\theta_{j, \rm R-L}(t)$ for a particular station. In a heterogeneous VLBI array, the type of fit depends on particular properties of each station, which may vary from a constant offset for multiple subsequent nights to a nonlinear trend varying on timescales of an hour. When available, we jointly analyze observations of multiple sources (e.g., scientific target and calibrators) when estimating source-independent station phase gain offsets over the course of a campaign. This makes the estimate robust against tuning to specific intrinsic source properties, such as contamination from a nonzero Stokes $V$ circular polarization component.

\begin{figure}[t]
    \centering
    \includegraphics[width=\linewidth]{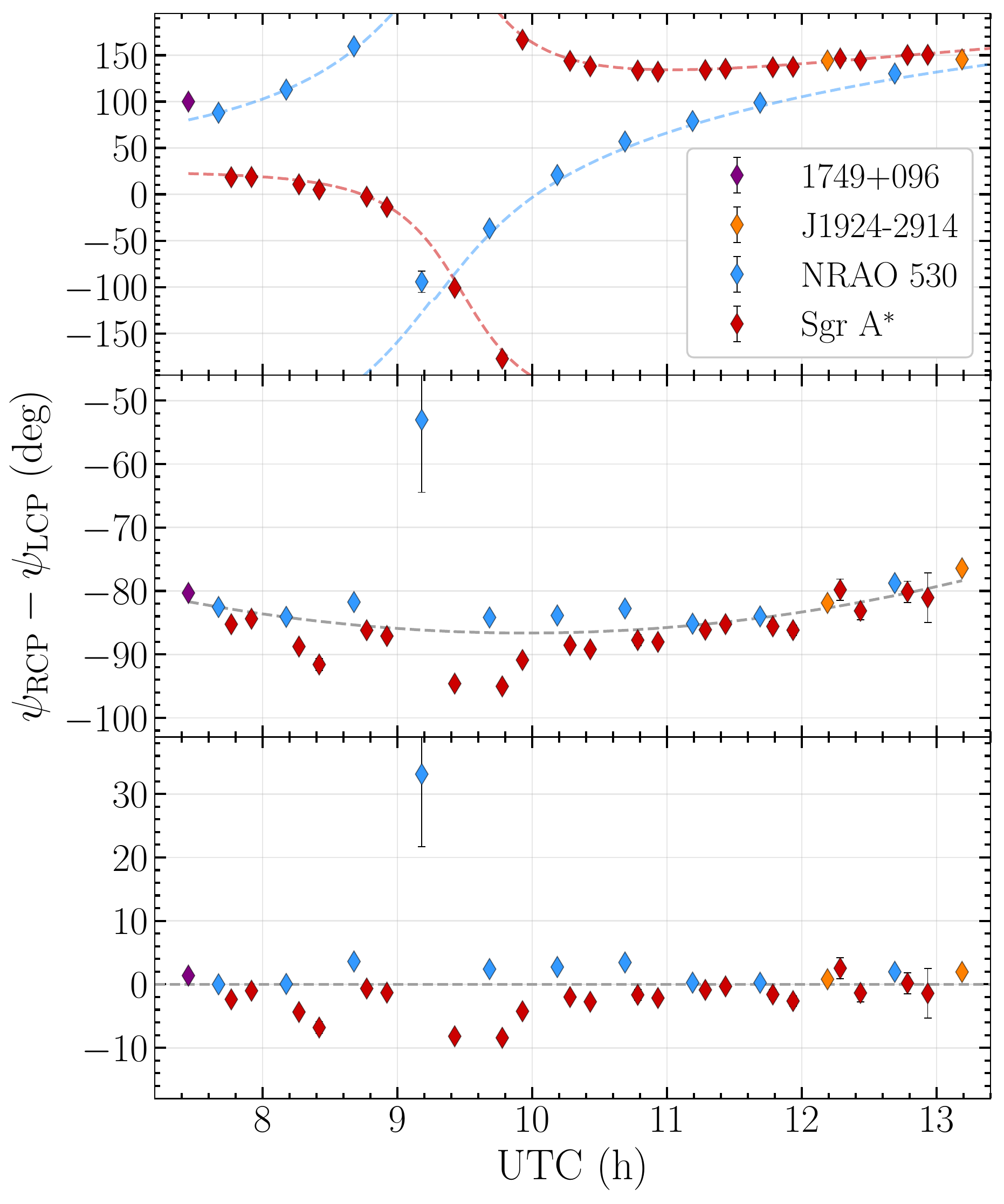}
    \caption{Steps of RCP--LCP phase offset calibration illustrated for the ALMA--GBT baseline in the GMVA+ALMA data set (\autoref{sect:dataset}). Top: phase offset after the fringe fitting step; middle: phase offset after parallactic angle correction; bottom: phase offset after global fitting of polarimetric gain ratio phases. The GBT phase gain is fitted as a~second-order polynomial to a~data set consisting of all four sources. In each case, the dashed line represents the full RCP--LCP phase model.
    }
    \label{fig:polcal}
\end{figure}

As an illustration, $\theta_{\rm R-L}(t)$ of the GBT, estimated from the GMVA+ALMA data set (see \autoref{sect:dataset}) is shown in the middle pandel of \autoref{fig:polcal} with a~dashed line. Here the offset was modeled as a~second-order polynomial and estimated using a data set consisting of four observed sources. For the phase gain offset calibration, one polarization component is chosen as a reference (LCP by default), and the other one is calibrated to match the first one. As an example, we have
\begin{align}
     &  V_{jk,\rm{RR}} \rightarrow V_{jk,\rm{RR}}  \exp \left[ -i(\theta_{j, \rm{R-L}} -\theta_{k, \rm{R-L}}) \right] , \nonumber \\
     &  V_{jk,\rm{LL}} \rightarrow V_{jk,\rm{LL}} \ .
\end{align}
A~final product of the polarimetric phase offset calibration is shown in the bottom panel of \autoref{fig:polcal}. In the top panel we also show the full polarimetric phase offset calibration model fit (field rotation plus gains) for \sgra\ and NRAO\,530 as dashed lines.

While the flux calibration, described in \autoref{sec:apriori}, is performed separately for different polarization products and is expected to account for a priori known differences in sensitivity between RCP and LCP at each site, the \texttt{eat} post-processing framework also offers an option to calibrate residual differences in RCP/LCP amplitude gain. If the option is selected, amplitude gain is estimated as an S/N-weighted median RCP/LCP amplitude ratio $|g_{j, \rm{R/L}}|$. Calibrating polarimetric amplitudes for the $jk$ baseline yields
\begin{align}
     &  V_{jk,\rm{RR}} \rightarrow V_{jk,\rm{RR}}  |g_{j,  \rm{ R/L} } |^{-1/2}|g^*_{k,  \rm{ R/L} }|^{-1/2} , \nonumber \\
     &  V_{jk,\rm{LL}} \rightarrow V_{jk,\rm{LL}}  |g_{j, \rm{ R/L}}|^{1/2} |g^*_{k, \rm{ R/L}}|^{1/2} \ .
\end{align}

The presented framework assumes a negligible influence of polarimetric leakage, the calibration of which is not yet a standard part of the EHT-HOPS data calibration pipeline. Proper calibration of leakage necessarily relies on the joint modeling of leakage terms, together with both polarized and unpolarized source structure \citep{Leppanen1995}.

\subsection{Network Amplitude Calibration}

The a~priori amplitude calibration of the EHT-HOPS pipeline (\autoref{sec:apriori}) can be improved by determining station-based corrections that produce visibility amplitude relationships that are expected from array redundancy. While array redundancy has regularly been used to improve calibration of connected element arrays, it has not been commonly used for VLBI \citep[see, e.g.,][]{Pearson_1984,cornwell_1989}. However, the EHT differs from standard VLBI arrays by including a number of colocated sites that introduce significant redundancy (e.g., three different facilities on Maunakea have participated in EHT observations: the Submillimeter Array [SMA], the James Clerk Maxwell Telescope [JCMT], and the Caltech Submillimeter Observatory), and this redundancy has been routinely utilized to derive amplitude calibration corrections \citep[e.g.,][]{Fish2011,Johnson2015,Lu2018}.  
We refer to the procedure of deriving these corrections as {\em network calibration}, and the EHT-HOPS implementation of network calibration is an extension of techniques developed in \citet{Johnson2015}. We now outline the assumptions, procedure, and limitations of network calibration.

\subsubsection{Network Calibration Assumptions}
\label{sec:netcal_assumptions}

Consider a VLBI array that contains one or more pairs of stations at a single geographic site (e.g., ALMA/APEX and SMA/JCMT for the EHT). Because intrasite baselines do not resolve the compact emission structure sampled on intersite baselines, they introduce consistency relationships that are weakly dependent on source structure. For example, letting $\mathcal{V}_x$ denote the ideal source visibility on the baseline $x$,
\begin{itemize}
    \item Intrasite visibilities should be equal to each other and to measurements of the total flux density $V_0$ made with connected element interferometers that sample the same angular scales, e.g., 
    \begin{align}
        \mathcal{V}_{\rm SMA-JCMT} = \mathcal{V}_{\rm ALMA-APEX} = V_0 \in \mathbb{R}^+.
    \end{align}
    \item Intersite visibilities to intrasite stations should be equal, e.g., 
    \begin{align}
        \mathcal{V}_{\rm LMT-SMA} = \mathcal{V}_{\rm LMT-JCMT}.
    \end{align}
\end{itemize}
Both of these properties follow from the assumption that intrasite baselines do not resolve the source; the first relationship integrates an additional measurement ($V_0$), which is routinely recorded in parallel with VLBI observations.

\subsubsection{Network Calibration Procedure}
\label{sec:netcal_procedure}

To motivate the network calibration procedure, we first consider visibility measurements with no thermal noise. Under the assumption that all systematic errors are station based, we can write a measured visibility $V_{ij}$ on a baseline between sites $i$ and $j$ as
\begin{align}
    V_{ij} = g_{\vphantom{j}i}^{\vphantom{\ast}} g_j^\ast \mathcal{V}_{ij},
\end{align}
where $g_i$ and $g_j$ are the station-based residual gains. 

Suppose that stations $i$ and $j$ are colocated, so that $\mathcal{V}_{ij} = V_0$. Knowledge of $V_0$ is not sufficient to determine $g_i$ and $g_j$, but measurements to a third site $k$ break the degeneracy:
\begin{align}
\label{eqn:netcal_ideal}
               |g_i| &= \sqrt{\left| \frac{V_{ij}}{V_0} \times \frac{V_{ik}}{V_{jk}} \right| },\\
     \nonumber |g_j| &= \sqrt{\left| \frac{V_{ij}}{V_0} \times \frac{V_{jk}}{V_{ik}} \right| }.
\end{align}
As these equations suggest, network calibration only determines the amplitudes of the gains of stations with colocated partners; it does not modify the gains of other stations or determine absolute phase corrections. In the limit of all stations having a colocated partner, network calibration yields absolute amplitude calibration for all stations.

Because the gains of an intrasite pair are fully determined by a third site, additional sites can be combined to reduce thermal uncertainties in the estimated gains. The EHT-HOPS pipeline uses all baselines simultaneously to solve for the set of unknown model visibilities $\mathcal{V}_{ij}$ and station gains $g_j$ by minimizing an associated $\chi^2$. Specifically, for each solution interval, we find the set of gains $\left\{ g_i \right\}$ and source visibilities $\left\{ \mathcal{V}_{ij} \right\}$ connecting each pair of sites by minimizing
 \begin{align}
     \chi^2 = \sum_{i<j} \frac{|g_{\vphantom{j}i}^{\vphantom{\ast}} g_j^* \mathcal{V}_{ij}-V_{ij}|^2}{\sigma_{ij}^2},
 \end{align}
 where $\sigma_{ij}$ is the thermal uncertainty on $V_{ij}$. In practice, the only gains that must be included are those of sites with intrasite partners; also, visibilities connecting two sites that each lack an intrasite partner can be excluded, as they provide no additional constraints for the network calibration. Thus, for $N$ sites of which $N_{\rm intra}$ have intrasite partners, network calibration requires solving for at most $N_{\rm intra}$ gains and $(N - N_{\rm intra}/2)(N-N_{\rm intra}/2-1)/2$ model visibilities. In 2017, the EHT had $N=8$ and $N_{\rm intra} = 4$, requiring solutions for at most 4 gains and 15 model visibilities in each solution interval.
 
We implemented this procedure for network calibration within the {\tt eht-imaging} library \citep{chael2016,Chael2018}. This calibration can be used for any VLBI array and only requires specifying the total source flux density $V_0$ and a maximum baseline separation for which a pair of sites is considered colocated (i.e., a threshold to define baseline lengths that do not resolve the observed source). 

\begin{figure*}[t]
  \includegraphics[width=\textwidth]{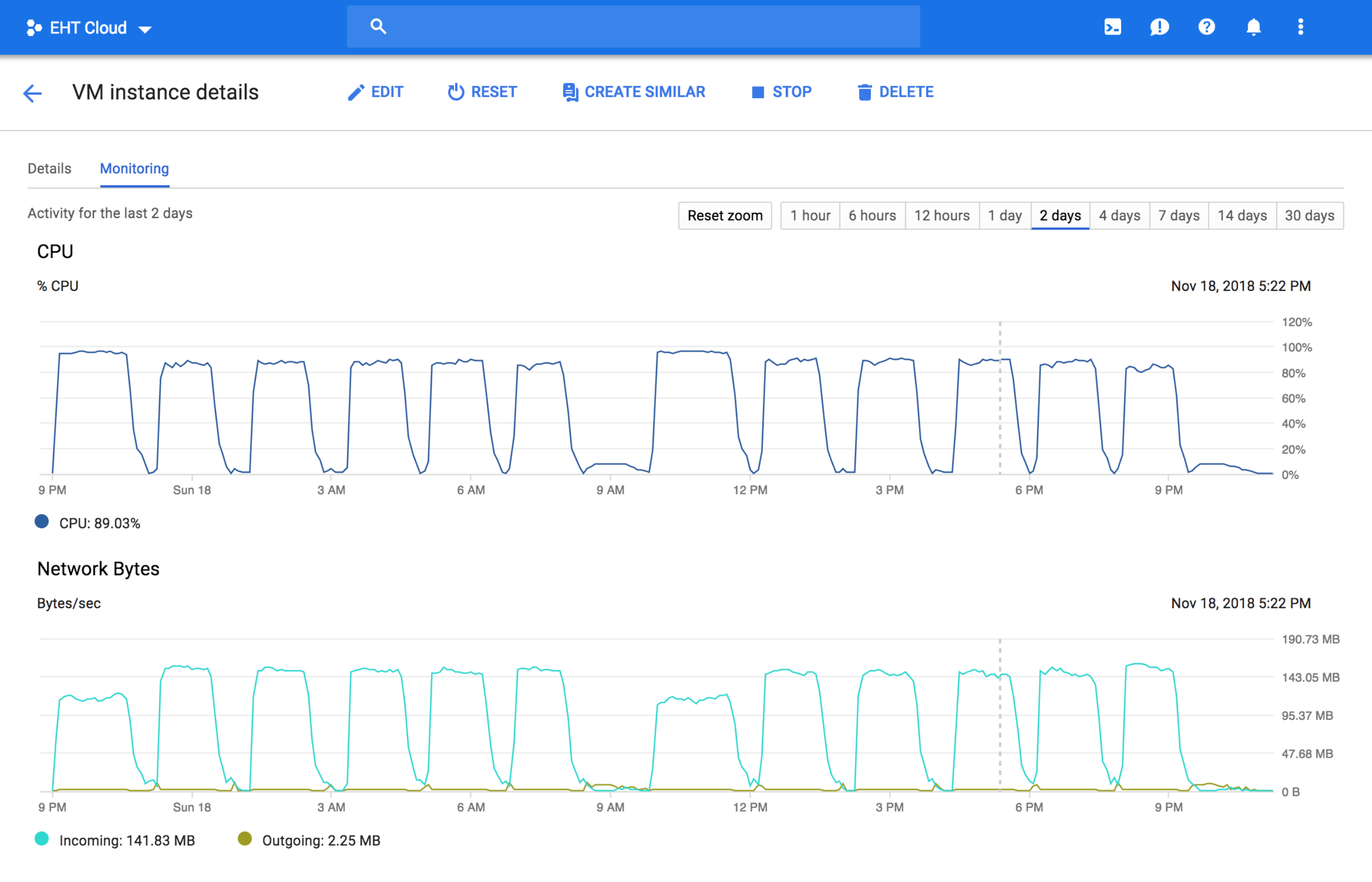}
  \caption{Screenshot of CPU and network usages from running the EHT-HOPS pipeline on a single 64-core virtual machine on Google Cloud Platform.
  The data volume processed is 1.7\,TB, covering two 2\,GHz bands from the EHT (up to eight stations, dual-polarization).
  Each pass processes one of the bands in about 12 hours, which includes one bootstrap stage with generic parameters and wide search windows, followed by the five fringe fitting stages of the EHT-HOPS pipeline (\autoref{fig:stages}).
  Each stage performs a fringe search over the complete data set.
  Because the \texttt{Mark4} formatted correlator input data are separated by scan, the processing at each stage can be naturally parallelized over multiple scans.}
  \label{fig:gcp}
\end{figure*}

\subsubsection{Network Calibration Error Budget}
\label{sec:netcal_errors}

The outcome of network calibration has both thermal errors from thermal noise on the input visibilities and systematic errors from broken assumptions in the network calibration procedure. We now briefly assess expected elements of this error budget.

Thermal errors are are straightforward to compute from analysis of the $\chi^2$ hypersurface explored in the minimization procedure discussed in \autoref{sec:netcal_procedure}. From \autoref{eqn:netcal_ideal}, it is clear that thermal errors have contributions from intrasite baselines and from intersite baselines; the latter typically have lower S/N because these baselines can heavily resolve the source, so they typically dominate the thermal error budget. For each pair of colocated sites $\{i,j\}$, the fractional uncertainty from thermal noise will be dominated by their strongest intersite baselines to another site $k$, with
\begin{align}
    \frac{\Delta g_i}{g_i}, \frac{\Delta g_j}{g_j} \sim \frac{1}{2} \sqrt{ \left(\frac{\sigma_{ij}}{|V_{ij}|} \right)^2 + \left(\frac{\sigma_{ik}}{|V_{ik}|} \right)^2 + \left(\frac{\sigma_{jk}}{|V_{jk}|} \right)^2 }.
\end{align}
For instance, in the case of ALMA and APEX, APEX will always have much lower S/N than ALMA and dominates the thermal uncertainties for the derived gains of both stations. If the maximum S/N from APEX to another site is 10 in the network calibration solution interval, then the network calibration will have a fractional uncertainty for $g_{\rm ALMA}$ and $g_{\rm APEX}$ of approximately 5\% from thermal noise. 

Systematic errors in the network calibration solution arise from incorrect or broken assumptions (see \autoref{sec:netcal_assumptions}). We now estimate the magnitude of three primary expected errors; additional sources of error (e.g., from baseline-dependent systematic errors) can be assessed using \autoref{eqn:netcal_ideal}.

\noindent {\em Incorrect assumed total flux density:} Errors in the assumed total flux density $V_0$ lead to a constant multiplicative factor for the derived gains. Suppose that $V_0 = \mathcal{V}_0 + \Delta V_0$, where $\mathcal{V}_0$ is the true total flux density. Then, $\Delta g / g \approx \sqrt{V_0/\mathcal{V}_0} \approx 1 + \frac{1}{2} \Delta V_0 / V_0$.

\noindent {\em Intrasite baselines partially resolve the source:} In practice, intrasite baselines may partially resolve the emission structure. In some cases, it may be possible to model this effect and correct for it (e.g., using an ALMA image to predict the flux density seen on the SMA--JCMT baseline). However, even in the limit of unmodeled losses, the measured flux density on a short baseline will differ from the true value by some amount $\Delta V_0$, and the error propagation is identical to the case of an incorrect assumed total flux density.

\noindent {\em Intersite baselines to intrasite partners are not equal:} Suppose that the intrasite stations are separated by a vector displacement $\mathbf{u}_{\rm intra}$, and let $\mathbf{u}_{\rm inter}$ denote an intersite baseline to a site that has an intrasite pair. In this case, network calibration relies on the approximation that the two intersite visibilities to the pair are approximately equal: $\mathcal{V}(\mathbf{u}_{\rm inter}) \approx \mathcal{V}(\mathbf{u}_{\rm inter} + \mathbf{u}_{\rm intra})$. By the van~Cittert--Zernike theorem, this condition can be expressed in terms of the sky brightness distribution \citep{tms}:
\begin{align}
\int d^2\mathbf{x}\, I(\mathbf{x}) e^{-2\pi i  \mathbf{x} \cdot \mathbf{u}_{\rm inter} } \hspace{-2.2cm}&\\
\nonumber &\approx \int d^2\mathbf{x}\, I(\mathbf{x}) e^{-2\pi i \mathbf{x} \cdot \left( \mathbf{u}_{\rm inter} + \mathbf{u}_{\rm intra} \right)  }\\
\nonumber &\approx \int d^2\mathbf{x}\, I(\mathbf{x}) e^{-2\pi i \mathbf{x} \cdot \mathbf{u}_{\rm inter} } \left( 1 - 2\pi i \mathbf{x} \cdot \mathbf{u}_{\rm intra} \right),
\end{align}
where $\mathbf{x}$ is an angular coordinate on the sky and $I(\mathbf{x})$ is the sky brightness distribution. The second approximation requires $u_{\rm intra} \ll 1/(2\pi x_{\rm max})$, where $x_{\rm max}$ is the maximum extent of nonzero source brightness. The fractional error in the first approximation is then of order $2\pi \mathbf{x} \cdot \mathbf{u}_{\rm intra}$. For the EHT, the longest intrasite baselines are shorter than intersite baselines by a factor of $u_{\rm inter}/u_{\rm intra} \gsim 10^3$. For sources that are weakly resolved on the shortest intersite baselines (i.e., $2\pi u_{\rm inter} x_{\rm max} \lsim 1$) the fractional error on a derived gain from breaking this assumption will then be $\Delta g/g \lsim 0.01$.

\Needspace*{4\baselineskip}
\section{Computing workflow}
\label{sec:compute}

The EHT-HOPS pipeline is designed to be automated and provide reproducible output.
The pipeline is conceptually structured in three layers:
1)~The software libraries/modules layer consists of the core software packages HOPS, \texttt{eat}, and \texttt{eht-imaging}.
2)~The driver scripts layer consists of \texttt{BASH} scripts for preparing input files, running programs from the software layer, creating logs and summary Jupyter notebooks, and cleaning up data products.
3)~The pipeline repository layer is made up of multiple directory structures that contain both configuration files for different processing stages (see \autoref{fig:stages}) and a master run script that enables running the full pipeline and packing output data products in a single step.
The software and driver script layers are generic and are suitable for being applied to different VLBI data sets.
Each pipeline repository, including summary notebook templates, is specific to a given production and to a specific data set.

To ensure reproducibility,
software libraries and module layers that are developed within the EHT, such as \texttt{eat}, are version controlled by \texttt{git} and publicly available on GitHub.
Furthermore, we use Docker, an operating-system-level virtualization software, to freeze the entire software environment, which includes many libraries and software packages distributed in binary format.
The recipes to build the Docker images, i.e., the Dockerfiles, are also version controlled and available on GitHub.

Although the entire pipeline can be run and debugged interactively on the native host operating system, production runs make use of Docker environments.
The associated hash-based Docker image identification numbers allow us to keep track of the exact versions of software, down to system libraries, and the specific image used for each production run is tagged along with its output.
This allows us to go back and repeat any previously tagged analysis.

The correlated data are separated by scan in relatively small ``type-1 corel'' individual files in the Mark4 file set.
When we run the EHT-HOPS pipeline, within each stage, all CPU cores on the (virtual) machine are made available to a single Docker container.
Inside the Docker container, we use GNU parallel to start multiple \texttt{fourfit} tasks, with one scan
mapped per task, to maximize CPU utilization.
When fringe fitting is done, we use the HOPS \texttt{alist} program to
reduce the fringe fitting results into a single summary text file. Further tools from \texttt{eat} process this output to generate HOPS calibration control codes for the next stage.
\autoref{fig:gcp} is a screenshot of CPU and network usage during a production run on a 64-core virtual machine on the Google Cloud Platform.
The periods of high CPU and network utilization correspond to the parallel \texttt{fourfit} tasks, while the periods of low utilization correspond to the \texttt{alist} and \texttt{eat} reduction tasks.
From the utilization cycles, it is easy to read off from \autoref{fig:gcp} that there are two passes of the data, one for each of the two 2\,GHz bands from the EHT.
Each pass includes one bootstrap stage with generic parameters and wide search windows, followed by the five fringe fitting stages with refined HOPS control files.

\section{Comparison to AIPS}
\label{sec:comparison}

\subsection{Data Set}
\label{sect:dataset}

The data set used for validation of the EHT-HOPS pipeline is the result of observations of \sgra\ on 2017 April 3 (project code MB007) with the GMVA, composed of the eight VLBA antennas operating at 86\,GHz, the Robert C. Byrd Green Bank Telescope (GBT), the Yebes 40m telescope (YS), the IRAM 30m telescope (PV), the Effelsberg 100\,m telescope (EB), and the ALMA phased array of 37 antennas. A total bandwidth of 512\,MHz (256\,MHz per polarization) was recorded at each station except YS, which recorded a single left circular polarization component. At correlation, the bandwidth per polarization was divided into four channels of 116 subchannels each. The observations included three calibrator sources: 1749$+$096, a bright quasar for bandpass and instrumental phase and delay calibration, and NRAO\,530 and J1924--2914, two quasars only $\sim$10$^\circ$ away from \sgra\ on the sky, for differential phase, delay, and rate calibration. Several of the VLBA stations (NL, OV, PT) observed in difficult weather conditions, such as frost, strong winds, or rain, leading to limited detections to those stations. Observations at PV suffered from phase instability and coherence losses in the signal chain, which led to poor-quality data and lower visibilities on its baselines. See \citet{Issaoun_2019} for further details on the observations. 

The data were reduced via the EHT-HOPS pipeline (\autoref{fig:stages}), with additional validation from the NRAO AIPS \citep{Greisen_2003}. Reduction in HOPS utilized ALMA, the most sensitive station in the array, as the reference antenna to derive stable instrumental phase bandpass and \mbox{RR--LL} delay relative to other stations (\autoref{sec:bandpass} and \ref{sec:rldelay}). Depending on S/N, either ALMA or GBT baselines were used to correct for intra-scan stochastic differential atmospheric phase, which varies on a timescale of a few seconds for this data set. The integration time for rapid phase corrections was determined automatically for each scan, taking into account the amount of random thermal variation and thus depending on S/N (\autoref{sec:adhoc}). In the final {\tt fourfit} stage, fringe solutions per scan were constrained to a single set of station-based delays and rates, or global fringe solutions, obtained from a least-squares solution to robust baseline detections (\autoref{sec:global}). A priori calibration was performed in post-processing, where all stations apart from YS, PV, and ALMA required an additional opacity correction to calibrate the visibility amplitudes (\autoref{sec:apriori}).

The AIPS reduction followed a classical procedure for low-frequency VLBI, with additional steps for fringe fitting refinement. After loading the data set into AIPS, during which digital sampler corrections are applied, we inspected the data interactively via the tasks {\tt BPEDT} and {\tt EDITA} and removed spurs in frequency domain accumulated bandpass tables and time domain amplitude plots. We then normalized the amplitudes via {\tt ACCOR} and applied field rotation angle corrections via {\tt VLBAPANG} (correcting for source parallactic angle and receiver mount type of each antenna) prior to fringe fitting. The standard instrumental phase calibration, with the station-based fringe fitter {\tt KRING}, corrects for experiment-wide correlator model phase and delay offsets using the full bandwidth and scan coherence. These solutions were derived using a scan on 1749+096, the brightest calibrator of the experiment, where 12 out of the 13 stations are present. A later scan on J1924--2914, the second-brightest calibrator, was used to derive solutions for the MK VLBA station, not present in the 1749$+$096 scans. ALMA was also used as the reference antenna for this processing. The instrumental phase calibration was applied to all scans before proceeding to finer fringe fitting, where either ALMA or GBT was used as a reference antenna, depending on the source. We solved for fringe rates and residual phase and delay offsets per channel, using full scan coherence, for each individual scan. We ran a third fringe fitting step to solve for stochastic atmospheric phase variations in time across the full bandwidth, with a fixed solution interval of 10\,s. A final fringe fitting step was used to solve for further scan-based residual delays and phases per channel to realign the channels. A priori calibration was performed with {\tt APCAL}, ignoring the opacity correction ({\tt DOFIT} $=-1$) for YS, ALMA, and PV.

\begin{table}[t]
\centering
\caption{Stokes $I$ detections in the GMVA+ALMA data.}
\begin{tabular}{ m{2cm}  m{1.5cm} m{1.5cm} } 
 \hline
 \hline 
 Source & AIPS & HOPS  \\ 
 \hline
 1749+096 & 120 & 123 \\ 
 J1924--2914 & 309 & 304  \\
 NRAO\,530 & 415 & 443 \\
 \sgra\ & 196 & 461 \\ 
 Total & 1040& 1331 \\
 \hline
\end{tabular}

\label{tab:detections}
\end{table}

\begin{figure}[t]
    \centering
    \includegraphics[width=\linewidth]{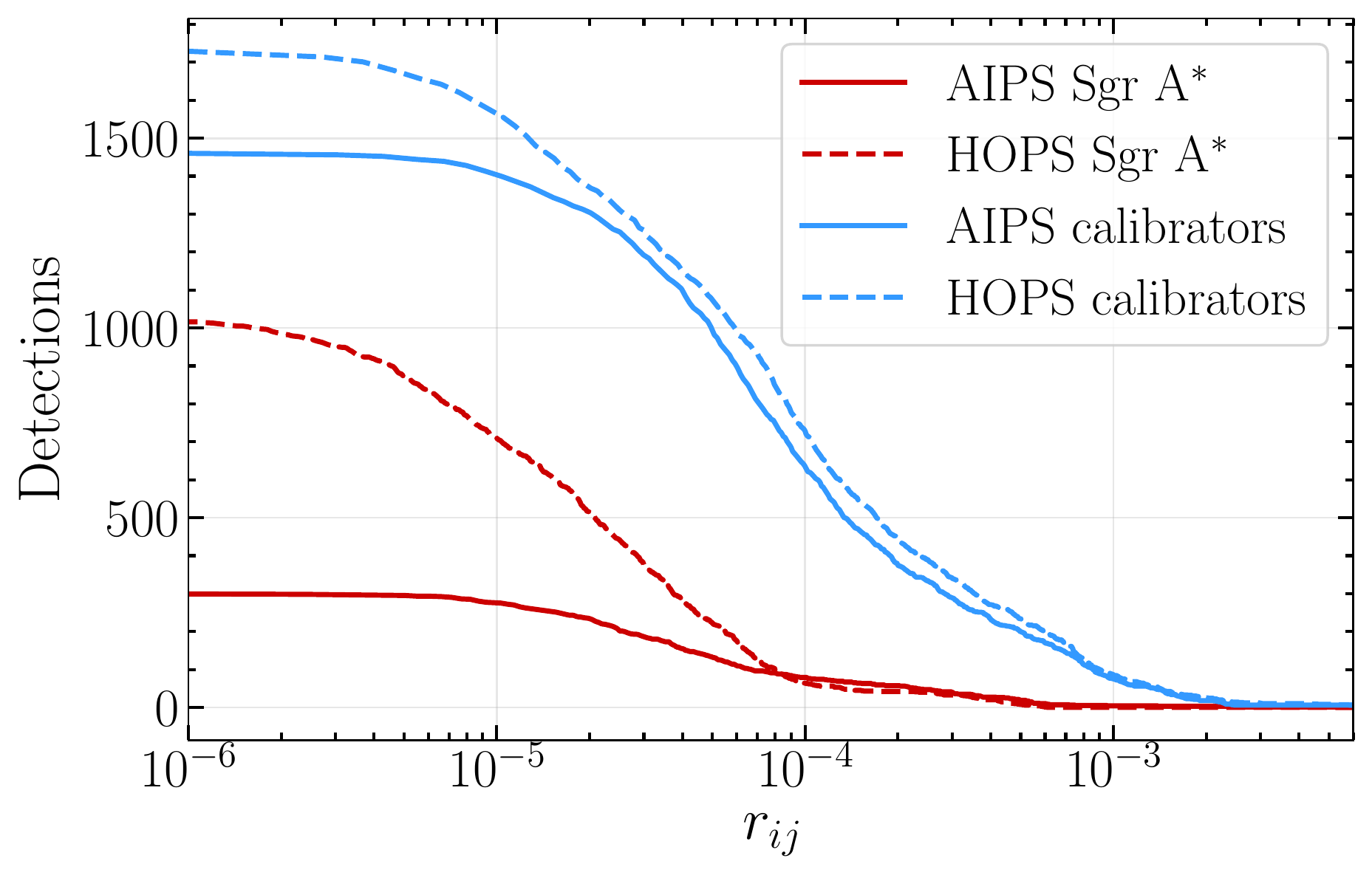}
    \caption{Comparison of cumulative histograms of correlation amplitude for HOPS and AIPS RR and LL detections. HOPS recovers a significant number of weak detections that are not present in the AIPS data product. Possible reasons for the differences in fringe recovery are discussed in \autoref{sec:pipelinecomparison}.
    }
    \label{fig:snr_hist}
\end{figure}

\begin{figure}[t]
    \centering
    \includegraphics[width=\linewidth]{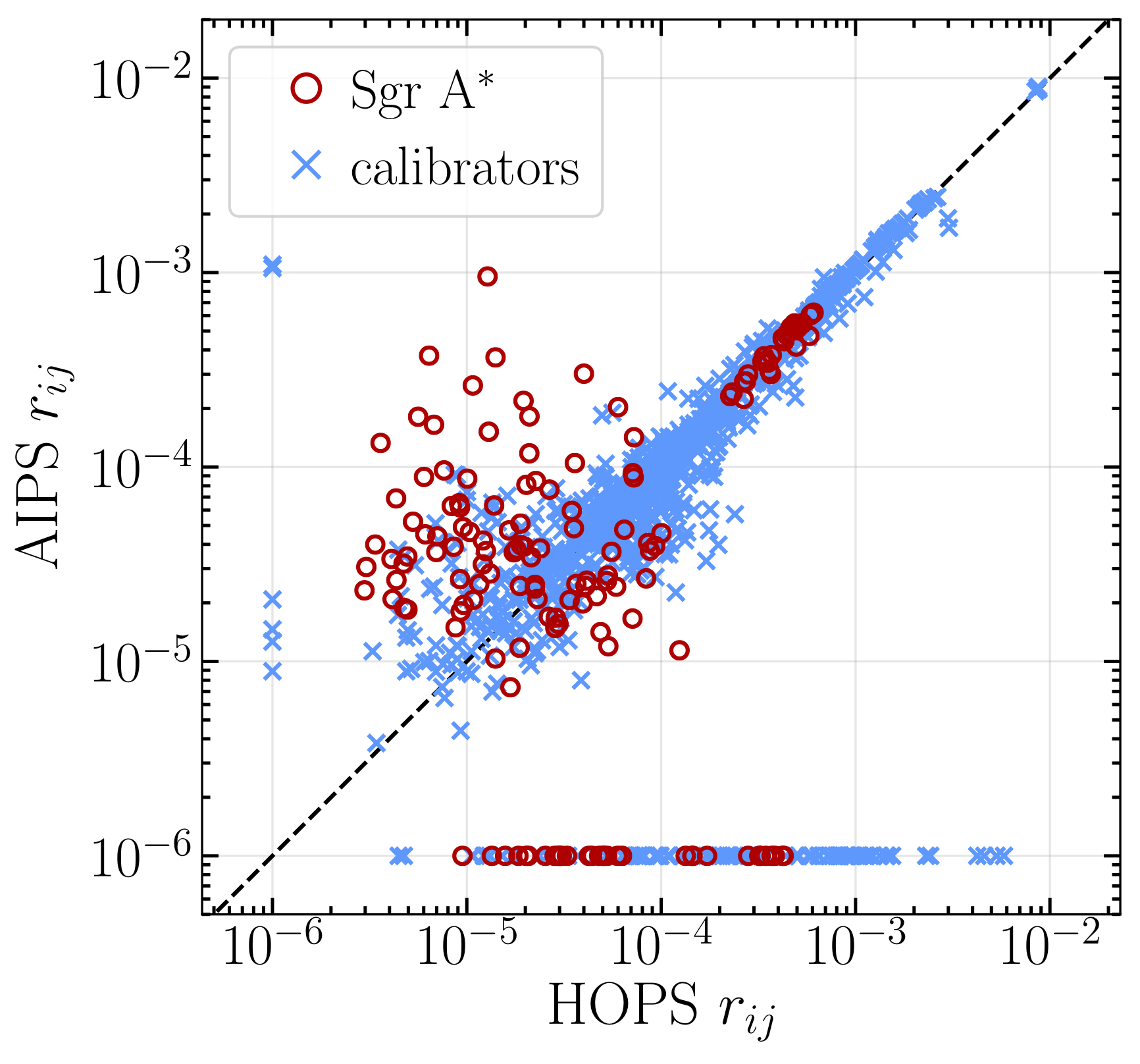}
    \caption{Scatter plot of correlation coefficient $r_{ij}$ magnitude in HOPS and AIPS data sets for RR and LL detections. The horizontal line at AIPS $r_{ij} = 10^{-6}$ corresponds to detections present exclusively in the HOPS data set.
    }
    \label{fig:rij_scatter}
\end{figure}

\begin{figure}[t]
    \centering
    \includegraphics[width=\linewidth]{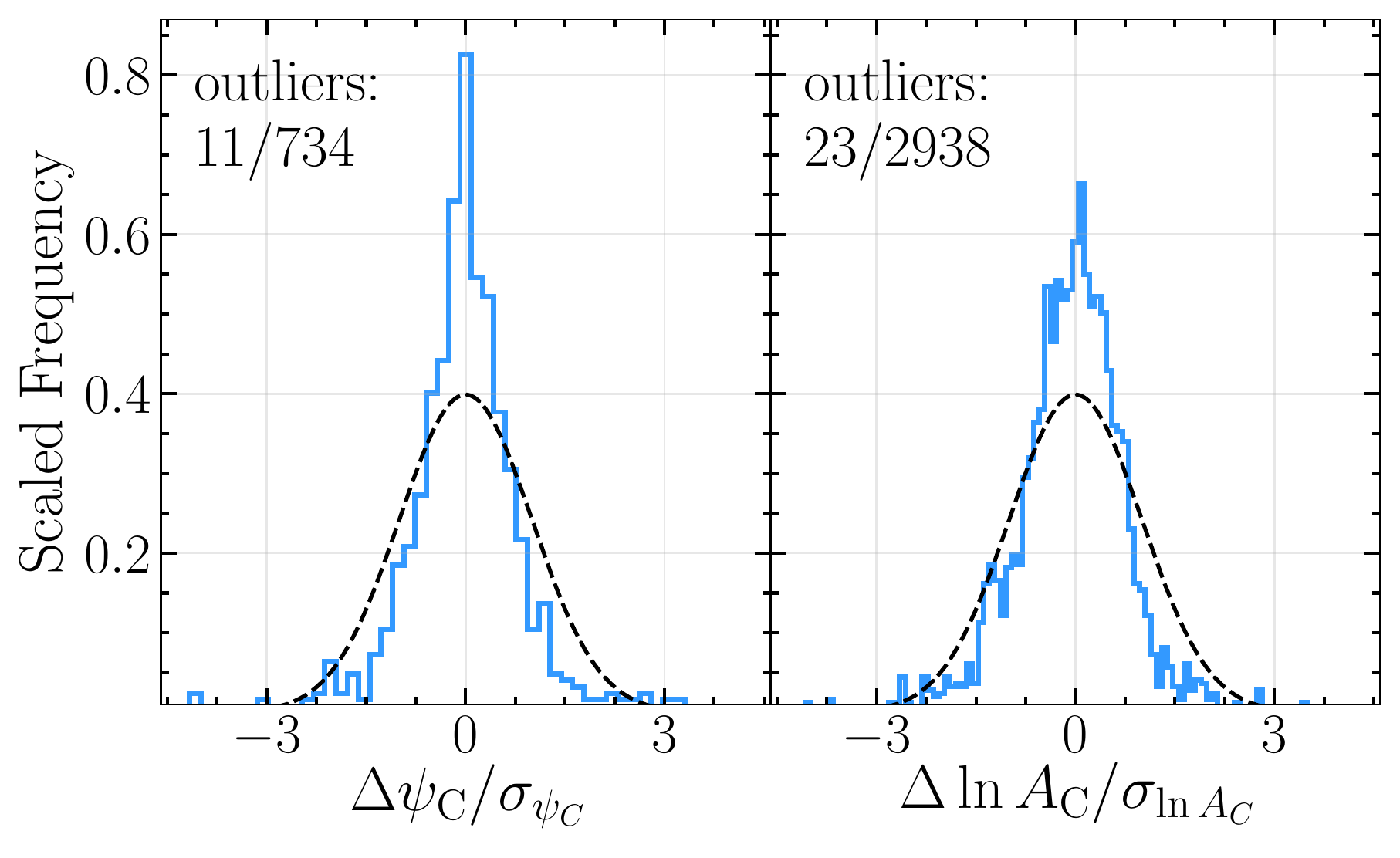}
    \caption{Histogram of differences HOPS - AIPS, normalized by the combined uncertainties of the pipelines for closure phases (left) and log closure amplitudes (right). The dashed line corresponds to a standard normal distribution.
    }
    \label{fig:cp_lca}
\end{figure}

\begin{figure}[t]
    \centering
    \includegraphics[width=\linewidth]{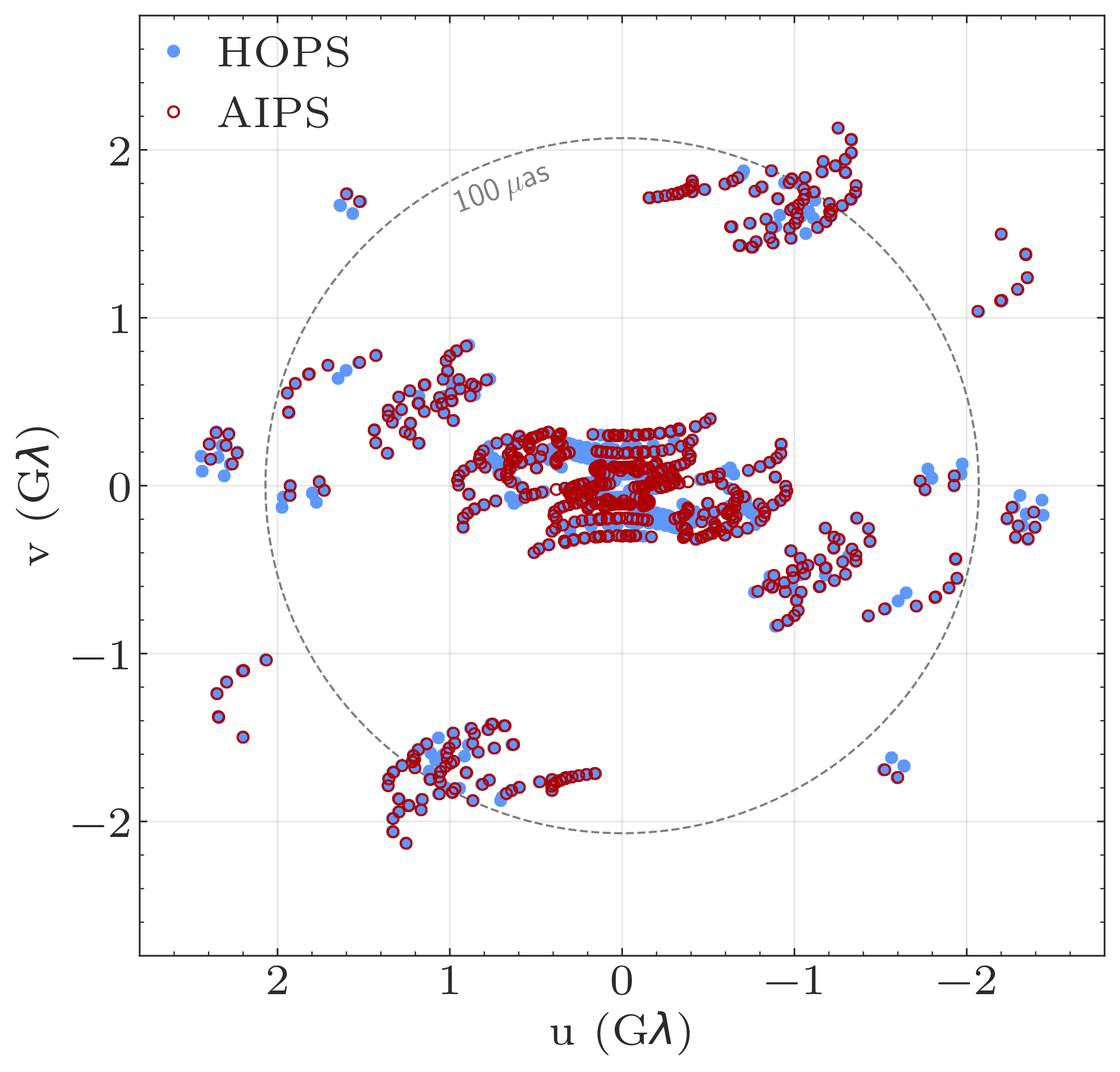}
    \caption{The $(u,v)$ GMVA+ALMA coverage of NRAO\,530 following data reduction. Each symbol denotes a scan-averaged measurement: filled blue circles are detections via the EHT-HOPS pipeline, and open red circles are detections via the AIPS processing.}
    \label{fig:nrao_uvcov}
\end{figure}

\begin{figure*}[t]
    \centering
    \includegraphics[width=0.497\linewidth]{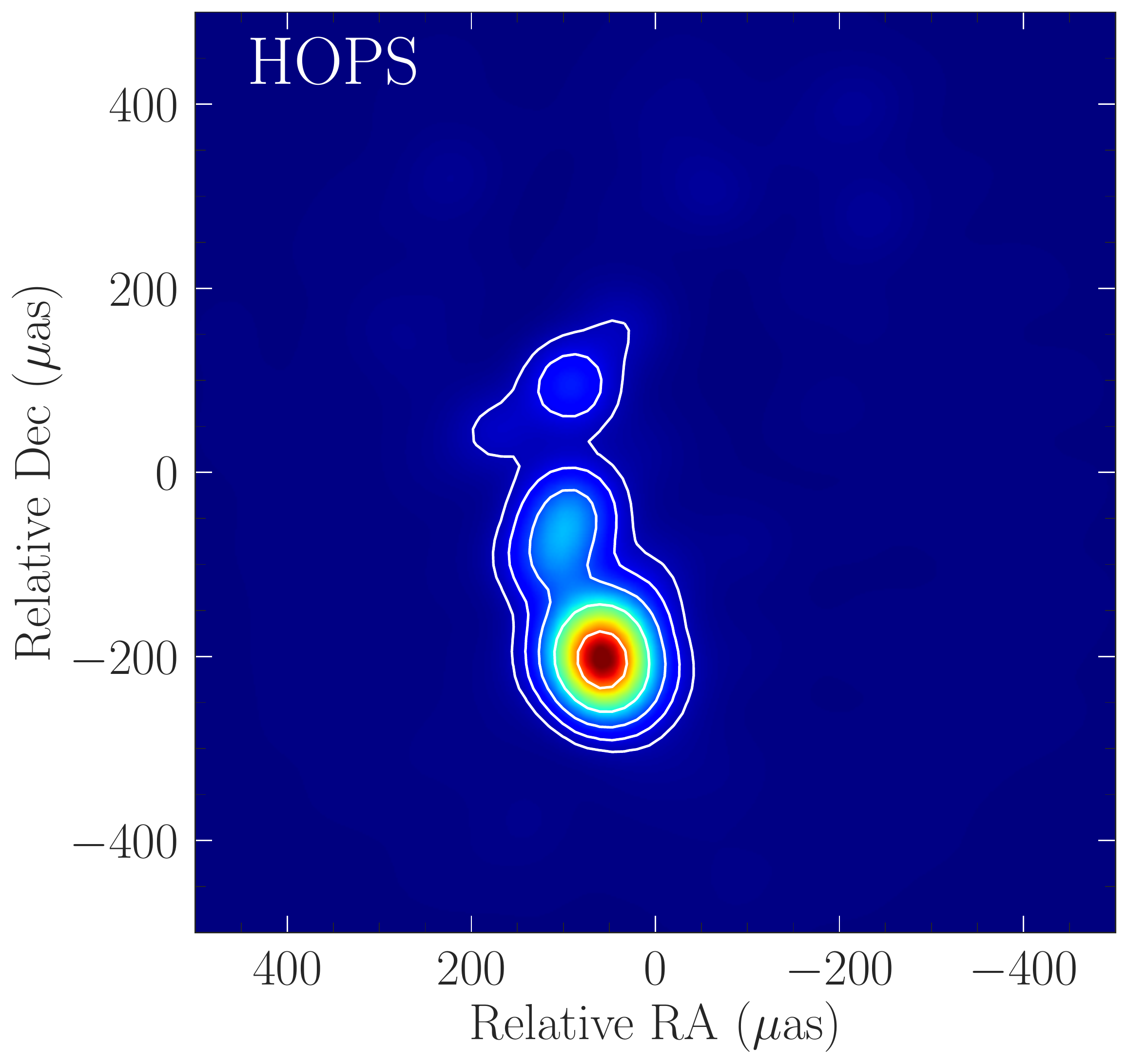}\hspace{-0.011\linewidth}
    \includegraphics[width=0.4926\linewidth]{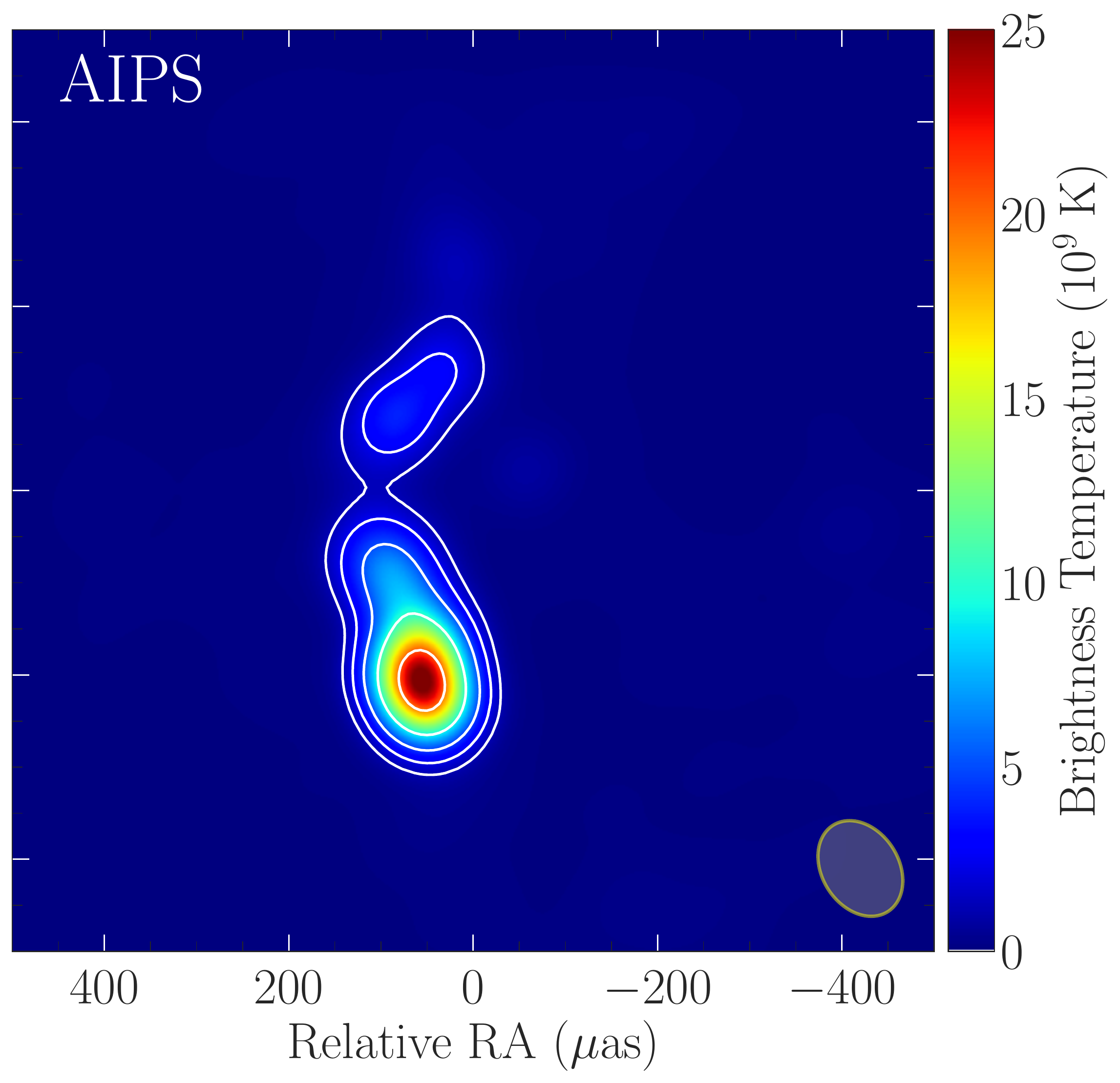}
    \caption{Closure-only images of NRAO\,530 as reduced by HOPS (left) and AIPS (right) and imaged with the {\tt eht-imaging} library \citep{Chael2018}. Total compact flux is determined by the analysis of ALMA interferometer data \citep{Goddi_2019}.
    The equal brightness temperature contour levels start from $1.25\times10^9$\,K and increase in factors of two. The observations have a uniform-weighted beam = $(111 \times 83)\,\mu$as, PA = $32^\circ$. }
    \label{fig:nrao_image}
\end{figure*}

\begin{figure}[t]
    \centering
    \includegraphics[width=\linewidth]{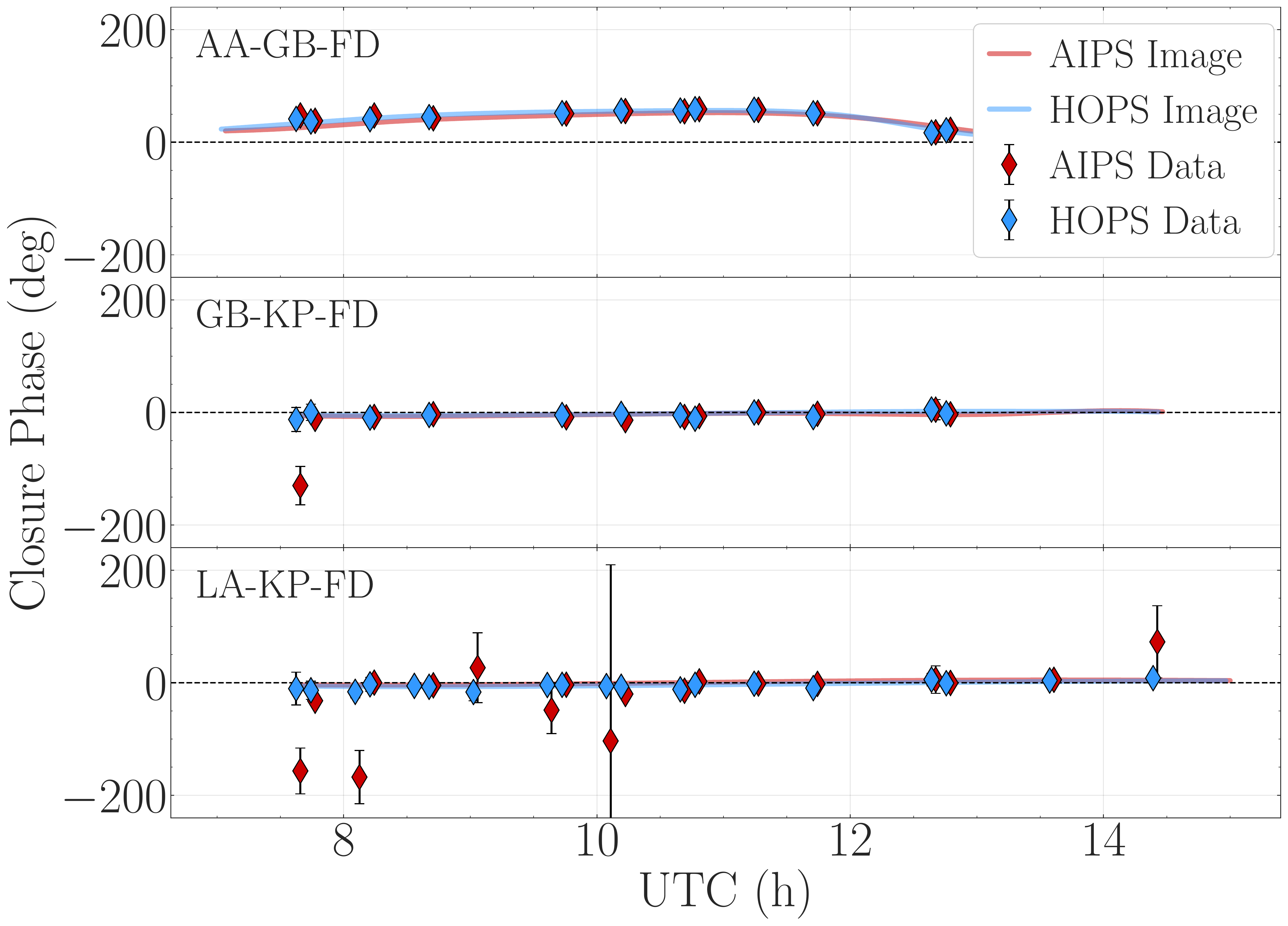}
    \caption{Scan-averaged closure phases for NRAO\,530 on three triangles (ALMA--GBT--FD, GBT--KP--FD, LA--KP--FD). Each diamond symbol denotes a scan-averaged measurement: blue diamonds are detections via the EHT-HOPS pipeline, and red diamonds are detections via the AIPS processing. The AIPS points are offset in time by $+$2\,min for clarity. Closure phase trends from the reconstructed images of the HOPS and AIPS data are shown as blue and red lines, respectively.
    }
    \label{fig:nrao_fit}
\end{figure}

\Needspace*{4\baselineskip}
\subsection{Pipeline Comparison}
\label{sec:pipelinecomparison}

We have performed a~comparative analysis of the GMVA+ALMA data set processed by the EHT-HOPS pipeline and a classic AIPS reduction. The EHT-HOPS pipeline has recovered a significantly larger number of detections, as summarized in \autoref{tab:detections}, as well as in \autoref{fig:snr_hist}. This likely reflects a more efficient use of free parameters for phase calibration in the EHT-HOPS pipeline. The EHT-HOPS pipeline calibration is driven by purpose-designed tasks targeting the characteristics of high-frequency VLBI data, while the AIPS processing relies on standard tasks available in the AIPS environment. A significant difference is in the handling of atmospheric phase, where the EHT-HOPS pipeline parameterizes phase variations as a smooth function using a flexible variability timescale that can accommodate the available S/N in the data (\autoref{sec:adhoc}). In our AIPS reduction, rapid phase variation is captured using a fixed 10\,s fringe solution interval, which may be too long (in the case of rapidly varying atmosphere for poor weather conditions or low elevation) or too short (in the case of low S/N). Benefits in sensitivity from the coherent Schwab-Cotton global fringe search in AIPS may not make up for the other inefficiencies owing to the arguments presented at the end of \autoref{sec:global}.

A~broad consistency between the pipelines can be seen in \autoref{fig:rij_scatter} for the common set of detections, showing the scatter plot of the correlation coefficient amplitude after the fringe fitting. While a certain amount of variation is seen in the lower-S/N part of the data set, particularly for \sgra, the high-S/N data show a high level of consistency between the two reductions. Additionally, we directly compare closure quantities in \autoref{fig:cp_lca}. Here we consider maximum sets of closure phases (left panel) and log closure amplitudes (right panel). We construct differences of scan-averaged RR and LL closure products matched between the pipelines and normalize them by the combined error for both pipelines. In the case of two independent measurements of the true underlying quantity, with normally distributed uncertainties, we would expect the result to be a~standard normal distribution, plotted with a dashed line for reference. The fact that the measured spread of the normalized differential quantity is smaller than that of a~standard normal distribution indicates that differences between the pipelines are of subthermal magnitude, even after full scan averaging. This is not surprising, as the pipelines are analyzing the same (not independent) thermal noise realizations. In \autoref{fig:cp_lca}, we also note a~small number of $3\,\sigma$ outliers.

In \autoref{fig:nrao_uvcov}, we illustrate the detections from both reductions on the $(u,v)$ coverage plot for the blazar NRAO\,530, in which the EHT-HOPS pipeline has recovered $\sim7$\% more detections. We proceed to an additional validation of the data sets via image reconstructions. We reconstruct images of NRA0\,530 with both the AIPS and EHT-HOPS data sets, constraining the total flux of the source from simultaneous ALMA interferometric measurements \citep{Goddi_2019}. For the imaging process, we make use of only closure quantities (amplitudes and phases), as the 13 stations of the GMVA+ALMA array and their coverage provide a large relative amount of closure information, independent of station gain errors \citep[e.g.,][]{tms}. We image NRAO\,530 using the {\tt eht-imaging} library \citep{chael2016}, following the closure imaging method of \citet{Chael2018}. The same script was used for both data sets, and the resulting images are shown in \autoref{fig:nrao_image}. The images show a high degree of consistency with each other, in addition to consistency in both morphology and jet direction with previous observations of NRAO\,530 in the literature \citep{Bower_1997,Bower_1998,Feng_2006,Chen_2010,Lu_2011b}.

In \autoref{fig:nrao_fit}, we inspect individual closure phases on three triangles of various sizes and orientations. The HOPS and AIPS closure phases are generally consistent, but the HOPS data set indicates smoother trends. The HOPS pipeline recovers zero closure phase more consistently on triangles that do not resolve the source, whereas AIPS has some difficulty owing to the lower S/N of the intra-VLBA detections (bottom panel of \autoref{fig:nrao_fit}). The closure phase trends derived from the two reconstructed images are also shown in \autoref{fig:nrao_fit}, and both images result in smooth trends in modeled closure phase that are similar to each other and either follow both data sets when the detections are well constrained or follow predominantly the HOPS detections when the AIPS detections result in different values.

While the calibrators are bright blazar sources typically reduced through classical AIPS procedures, the case of \sgra\ presents added difficulty to the calibration process. In particular, the source is subject to interstellar scattering in our line of sight, causing scatter broadening predominantly in the east--west direction, where a large majority of GMVA baselines lie \citep{Davies_1976,vanLangevelde_1992,Frail_1994,Bower_2004,Bower_2006,Shen_2005,Johnson_2018,Psaltis_2018}. Additionally, \sgra\ was $\sim$2\,Jy in 2017 at 3.5\,mm, at the lower end of its typical flux density range at this wavelength, and most stations of the GMVA, in the northern hemisphere, observe \sgra\ at very low elevations and thus through a large air mass that lowers the chance for strong detections. These conditions add difficulty to a classical AIPS processing, which does not fare as well for \sgra\ fringe fitting as the EHT-HOPS pipeline. Due to the clear difference in performance, as shown in \autoref{tab:detections} and \autoref{fig:snr_hist}, the EHT-HOPS pipeline processing was chosen to derive subsequent scientific results on \sgra, presented in \citet{Issaoun_2019}.

\Needspace*{4\baselineskip}
\section{Summary}

We have developed an automated calibration and reduction pipeline for high-frequency VLBI data, suitable for processing data from the GMVA (at 86\,GHz) and the EHT (at 230\,GHz). The pipeline is structured around the Haystack Observatory Post-processing System (HOPS), which was originally designed for precision geodetic analysis but has also been widely used for the processing of early data from the EHT. The new \mbox{EHT-HOPS} pipeline was targeted to meet the needs of the developing EHT and GMVA arrays. Specifically, it leverages high-sensitivity anchor stations, such as ALMA acting as a phased array, in order to phase-stabilize the network to atmospheric turbulence. The pipeline also provides reduced data that are phase calibrated to a global fringe solution in a standard \texttt{UVFITS} data format. This allows the HOPS output to be analyzed using a wide variety of downstream tools for VLBI data characterization, imaging, and modeling.

The EHT-HOPS pipeline was successfully used for the analysis of VLBI data taken at 86\,GHz on \sgra\ and associated calibration sources, using the GMVA joined by the ALMA phased array. The scientific analysis of the data was presented in \citet{Issaoun_2019}, leading to the first VLBI images of the intrinsic compact radio core of \sgra and the first VLBI results with ALMA. In this work we have used data from the observations to illustrate the calibration process and have compared the output from the EHT-HOPS pipeline with a classical data reduction through AIPS.
The EHT-HOPS pipeline was also applied as one of three independent reduction pipelines to the 230\,GHz (1.3\,mm) observations from the EHT 2017 April campaign \citep{PaperII,PaperIII}, where it showed a high degree of consistency with parallel reductions in AIPS (using a similar reduction to that presented in this work) and CASA \citep[using rPICARD;][] {janssen2019}. The scientific analysis of the EHT 2017 data set resulted in the first images and characterization of a black hole ``shadow" at the center of the radio galaxy M87 \citep{PaperI,PaperIV,PaperV,PaperVI}.

The current implementation of the pipeline addresses the need for rapid phase calibration at high observing frequencies and focuses on the robust detection of correlated fringes for the newly expanded VLBI networks. Future developments to the \texttt{UVFITS} post-processing tool set will support amplitude bandpass corrections and polarization leakage corrections, to reduce nonclosing baseline systematic errors and to provide the calibration necessary for polarization analysis.

\begin{acknowledgements}

We thank Thomas Krichbaum and Iv\'an Mart\'i-Vidal for helpful comments and discussion, as well as the EHT Collaboration, particularly members of the Correlation and the Calibration \& Error Analysis working groups.
We thank the National Science Foundation (AST-1126433, AST-1614868, AST-1716536) and the Gordon and Betty Moore Foundation (GBMF-5278) for financial support leading to this work. This work was supported in part by the Black Hole Initiative at Harvard University, which is supported by a grant from the John Templeton Foundation. The work was also supported by the ERC Synergy Grant ``BlackHoleCam: Imaging the Event Horizon of Black Holes,'' Grant 610058. This paper makes use of the following ALMA data: ADS/JAO.ALMA2016.1.00413.V. ALMA is a partnership of ESO (representing its member states), NSF (USA), and NINS (Japan), together with NRC (Canada), MOST and ASIAA (Taiwan), and KASI (Republic of Korea), in cooperation with the Republic of Chile. The Joint ALMA Observatory is operated by ESO, AUI/NRAO, and NAOJ. This research has made use of data obtained with the Global Millimeter VLBI Array (GMVA), which consists of telescopes operated by the Max-Planck-Institut f{\"u}r Radioastronomie (MPIfR), IRAM, Onsala, Metsahovi, Yebes, the Korean VLBI Network, the Green Bank Observatory, and the Very Long Baseline Array (VLBA). The VLBA is a facility of the National Science Foundation operated under cooperative agreement by Associated Universities, Inc. The data were correlated at the correlator of the MPIfR in Bonn, Germany. This work is partly based on observations with the 100\,m telescope of the MPIfR at Effelsberg. This work made use of the Swinburne University of Technology software correlator \citep{deller2011}, developed as part of the Australian Major National Research Facilities Programme and operated under licence.
\end{acknowledgements}

\facilities{GMVA (VLBA, Effelsberg, IRAM:30m, Yebes), GBT, ALMA}

\software{HOPS \citep{Whitney_2004}, AIPS \citep{Greisen_2003}, GNU Parallel \citep{GNU}, eht-imaging \citep{chael2016,Chael2018}, Numpy \citep{numpy_2011}, Scipy \citep{scipy}, Pandas \citep{pandas}, Astropy \citep{astropy_2013,astropy_2018}, Jupyter \citep{jupyter}, Matplotlib \citep{Hunter_2007}.}

\bibliography{main.bib}{}
\bibliographystyle{yahapj.bst}

\end{document}